\documentclass[twocolumn]{aastex62}
\usepackage{graphicx}   
\usepackage{float}      
\usepackage{subfigure}  
\usepackage{multirow}
\usepackage{xcolor}

\newcommand{\kms}{km\,s$^{-1}$}

\newcommand{\teff}{T_{\rm eff}}

\newcommand{\vt}{\xi_t}

\def\gtsim {>\kern-0.9em\lower1.1ex\hbox{$\sim$}~}
\def\ltsim {<\kern-0.9em\lower1.1ex\hbox{$\sim$}~}

\accepted{\today}

\submitjournal{\apj}

    \shorttitle{Yan et al}
\shortauthors{Yan et al.}

\begin{document}

\title{A Systematic NLTE Study of Very Metal-Poor Stars with Metallicity Down to $-4.3$ dex. \\
II. Lithium Abundance and New Insight to the Lithium Plateau}

\author[0000-0002-8609-3599]{Hong-Liang Yan}
\affiliation{National Astronomical Observatories, Chinese Academy of Sciences, Beijing 100101, China}
\affiliation{Institute for Frontiers in Astronomy and Astrophysics, Beijing Normal University, Beijing 102206, China}
\affiliation{School of Astronomy and Space Science, University of Chinese Academy of Sciences, Beijing 100049, China}

\author[0009-0000-2934-4067]{Jinxiao Qin}
\affiliation{Department of Physics, Hebei Normal University, Shijiazhuang 050024, People's Republic of China}
\affiliation{National Astronomical Observatories, Chinese Academy of Sciences, Beijing 100101, China}
\affiliation{Institute for Frontiers in Astronomy and Astrophysics, Beijing Normal University, Beijing 102206, China}

\author[0000-0001-5193-1727]{Shuai Liu} 
\affiliation{National Astronomical Observatories, Chinese Academy of Sciences, Beijing 100101, China}

\author[0000-0002-1619-1660]{Zeming Zhou} 
\affiliation{National Astronomical Observatories, Chinese Academy of Sciences, Beijing 100101, China}

\author[0000-0002-8980-945X]{Gang Zhao} 
\affiliation{National Astronomical Observatories, Chinese Academy of Sciences, Beijing 100101, China}
\affiliation{School of Astronomy and Space Science, University of Chinese Academy of Sciences, Beijing 100049, China}

\author[0000-0002-0349-7839]{Jianrong Shi} 
\affiliation{National Astronomical Observatories, Chinese Academy of Sciences, Beijing 100101, China}
\affiliation{School of Astronomy and Space Science, University of Chinese Academy of Sciences, Beijing 100049, China}

\author[0000-0002-8709-4665]{Sofya Alexeeva} 
\affiliation{National Astronomical Observatories, Chinese Academy of Sciences, Beijing 100101, China}

\author[0000-0002-7727-1699]{Huawei Zhang} 
\affiliation{Department of Astronomy, School of Physics, Peking University, Beijing 100871, China}
\affiliation{Kavli Institute of Astronomy and Astrophysics, Peking University, Beijing 100871, China}

\author[0000-0002-0389-9264]{Haining Li} 
\affiliation{National Astronomical Observatories, Chinese Academy of Sciences, Beijing 100101, China}

\author[0009-0008-2988-2680]{Huiling Chen}
\affiliation{Department of Astronomy, School of Physics, Peking University, Beijing 100871, China}
\affiliation{Kavli Institute of Astronomy and Astrophysics, Peking University, Beijing 100871, China}

\author[0000-0002-3395-6178]{Junbo Zhang} 
\affiliation{National Astronomical Observatories, Chinese Academy of Sciences, Beijing 100101, China}

\author[0000-0003-4445-6504]{Yufu Shen} 
\affiliation{Changchun Observatory, National Astronomical Observatories, Chinese Academy of Sciences, Jingyuetan National Scenic Area, Changchun 130117, China}

\author[0000-0002-8975-6829]{Wako Aoki} 
\affiliation{National Astronomical Observatory of Japan, 2-21-1 Osawa, Mitaka, Tokyo 181-8588, Japan}
\affiliation{Astronomical Science Program, The Graduate University for Advanced Studies, SOKENDAI, 2-21-1 Osawa, Mitaka, Tokyo 181-8588, Japan}

\author[0000-0002-8077-4617]{Tadafumi Matsuno} 
\affiliation{Astronomisches Rechen-Institut, Zentrum f\"ur Astronomie der Universit\"at Heidelberg, M\"onchhofstra{\ss}e 12-14, 69120 Heidelberg, Germany}

\author[0000-0003-2868-8276]{Jingkun Zhao}
\affiliation{National Astronomical Observatories, Chinese Academy of Sciences, Beijing 100101, China}

\correspondingauthor{Hong-Liang Yan and Gang Zhao}
\email{hlyan@nao.cas.cn; gzhao@nao.cas.cn}

\begin{abstract} \label{section:abstract}

Metal-poor stars are crucially important for understanding the early Galaxy, first stars, and the Universe. In this series of papers, we present a homogeneous non-local thermodynamic equilibrium (NLTE) abundances analysis of 12 elements for 103 very/extremely metal-poor (VMP/EMP) stars with metallicity down to $-4.3$ dex.
The sample was selected from the LAMOST survey and observed by the high-resolution spectroscopy of Subaru. 
In this paper, we present the NLTE abundances and evolution of lithium in these stars. 
We report different lithium behaviors corresponding to different evolutionary stages and their signatures: 
1) The Spite Plateau shows a slightly positive slope, indicating increasing lithium abundance with increasing metallicity. Most significantly, it appears to extend to lower metallicities as previously suggested, calling into question the reality of the so-called 'meltdown' at low metallicity; 
2) We confirm a lithium plateau for lower red giant branch (LRGB) stars with A(Li) $= 1.13$\,dex in our sample, while lithium abundance drops rapidly to A(Li)$<0.5$ as stars continue to evolve to higher stage. 
3) We identify four Li-rich stars in our sample across different evolutionary stages, showing complex and multiple lithium production mechanisms in VMP/EMP stars.
These findings suggest that early Galactic lithium enrichment results from a complex interplay between depletion and production processes.

\end{abstract}


\section{Introduction} \label{section:introduction}

Metal-poor stars\footnote{Based on their iron deficiency, metal-poor stars are commonly classified as metal-poor, very metal-poor (VMP), extremely metal-poor (EMP), and ultra metal-poor (UMP) stars, etc.. 
For convenience, we refer to our sample stars as VMP stars, although their metallicities span the range from metal-poor to extremely metal-poor.} are crucial for understanding the earliest epochs of the Galaxy and the Universe.
Their chemical composition offers the signatures of early generation (i.e. Population III) stars, first supernovae, and interstellar medium of the proto-Galaxy \citep{2005ARA&A..43..531B, 2013ARA&A..51..457N, 2015ARA&A..53..631F, 2019MNRAS.482.1204H}. 
Moreover, their spatial and kinematic properties provide insight into the hierarchical assembly of the Milky Way halo, revealing remnants of accreted dwarf galaxies and ancient stellar streams \citep{2018Natur.563...85H, 2018MNRAS.475.1537M}. 

Systematic efforts over time have yielded increasingly robust abundance determinations for metal-poor stars \citep[e.g.,][]{2004A&A...416.1117C, 2013ApJ...778...56C, 2014ApJ...787..162H, 2015ApJ...808..148S, 2016ApJ...833..225Z, 2020A&A...642A..25F, 2022ApJ...931..146A, 2022ApJ...931..147L}, and a comprehensive overview of these advances and additional references can be found in \citet{2025A&ARv..33....2B}. Together with stars in the metal-rich end, stellar abundances across a wide metallicity range are being investigated and compared with the Galactic chemical evolution (GCE) models \citep[e.g.,][]{2010A&A...522A..32R, 2020ApJ...895..138K}, providing insights into the roles of different nucleosynthesis channels \citep{1995ApJS..101..181W}, and challenging monolithic scenarios as well \citep[e.g.,][]{2018A&A...612A..65B, 2018PASJ...70...94A}. 

Among all the elements studied in metal-poor stars, lithium is uniquely important. 
$^7$Li was initially synthesized in Big Bang Nucleosynthesis (BBN). According to standard BBN theory, the primordial lithium abundance is determined by the baryon-to-photon ratio, $\eta$. Since the photon density is precisely determined by the present-day cosmic microwave background (CMB) radiation, the critical parameter is the baryon density, which can be inferred not only from CMB anisotropies \citep[e.g., WMAP and Planck, $\Omega_b h^2 = 0.02225 \pm 0.00016$,][]{2016A&A...596A.109P}, but also independently from measurements of primordial deuterium \citep{2018ApJ...855..102C}, both of which provide consistent constraints. 
This theoretical expectation, however, stands in persistent tension with the so-called `Li Plateau' or `Spite Plateau' \citep{1982A&A...115..357S, 1982Natur.297..483S}, which refers to a roughly constant lithium abundance observed in warm metal-poor ([Fe/H] $< -1.5$) dwarf stars, regardless of metallicity or temperature.
The Spite Plateau was initially interpreted as direct evidence of primordial lithium. However, the lithium abundance corresponds to the Spite Plateau lies significantly below the BBN predictions \citep{2004ApJ...600..544C, 2016RvMP...88a5004C} by a factor of 2$\sim$3 \citep{2010A&A...522A..26S, 2018A&A...612A..65B}. 
This discrepancy is referred as the `cosmological lithium problem'. 
The physical reason for this discrepancy remains debated, with proposed solutions ranging from stellar lithium depletion to revisions of standard BBN. 
Early studies dismissed significant depletion due to the plateau’s uniformity, however, lithium abundances inferred from CMB measurements now require astrophysical explanations, such as pre-main-sequence destruction \citep{2015MNRAS.452.3256F} or inhomogeneous baryon density fluctuations \citep{2009PhR...472....1I}. Recent studies report $A(\mathrm{Li}) \sim 2.2$ in low-metallicity gas clouds, suggesting that stellar depletion may not fully resolve the cosmological lithium problem given that the Li abundance of gas-phase cloud is not affected by this process \citep{2024A&A...690A..38M}.

For stars with metallicity below $\sim -2.5$\,dex, the constancy of the Spite Plateau has been the subject of ongoing debate. 
Previous works have found that there is a `meltdown' or `break' of the Spite Plateau. 
For example, \cite{1999ApJ...523..654R} pointed out the slope and discussed the possible increase of lithium with metallicity even in the metal-poor range, assuming lower primordial abundance. 
\cite{1997MNRAS.285..847B} attributed the discrepancies to the biases in the temperature scale. \cite{2010A&A...522A..26S} showed that the scatter in lithium abundance increases as metallicity decreases for warm halo stars in the metallicity range between $\sim -2.6$ and $\sim -3.5$\,dex. 
\cite{2009ApJ...698.1803A} reported that the average lithium abundances for stars with metallicity lower than $-3.0$ is lower by $0.2$ dex than that of stars with higher metallicity.
\cite{2010A&A...515L...3M} suggested two levels of the plateau for metallicity ranges higher and lower than [Fe/H]=$-2.5$. \cite{2017AJ....154...52M} argued that stars might show some sort of plateau at [Fe/H] $< -3.5$.
\cite{2022ApJ...931..147L} found that the lithium plateau could be seen in turnoff stars and the average Li abundance is clearly lower at lower metallicity.
The meltdown of the Spite plateau may indicate a transition between primordial and stellar-produced lithium, but its exact cause remains debated. 

In addition, recent works \citep[e.g.,][]{2012MNRAS.419.2195M, 2022A&A...661A.153M, 2022ApJ...931..147L} suggested that there is a thin lithium plateau in the metal-poor low red giant branch stars, which indicate that the signal of plateau can be preserved during stellar evolution, and thus the signature (if there is any) of lithium abundance in evolved stars could be also used as constrains to the primordial value.

Non-local thermodynamic equilibrium (NLTE) effects critically influence abundance determinations, particularly for ionization-sensitive elements \citep[e.g.,][]{1998A&A...333..219Z, 2000A&A...362.1077Z, 2004A&A...413.1045G, 2005ARA&A..43..481A, 2006A&A...457..645Z, 2008A&A...481..489Z, 2008A&A...486..303S, 2014AstL...40..406A, 2015ApJ...802...36Y, 2016ApJ...833..225Z, 2017A&A...608A..89M, 2018ApJ...866..153A, 2023ApJ...957...10A}. 
From another perspective, in studies of stars with metallicities as low as one-thousandth or even one ten-thousandth of the solar value, LTE assumptions may underestimate effective temperatures ($\teff$) by hundreds of Kelvins or overestimate [Fe/H] due to misinterpretations of Fe\,I/Fe\,II ionization balances \citep{2017A&A...604A.129M}. 
Previous works reported that the abundance discrepancy between Fe\,I and Fe\,II can be significant, reaching $\sim 0.4$--$0.8$ dex under LTE assumptions \citep[e.g.,][]{2012MNRAS.427...50L,2017ApJ...847..142E,2017A&A...597A...6N,2016MNRAS.463.1518A}. 
It should be noted that recent work reported consistent Fe\,I and Fe\,II abundances by LTE analysis within uncertainties for HE\,0107$-$5240 \citep{2025A&A...704A.238C}, one of the most metal-poor stars \citep{2002Natur.419..904C}, suggesting that the NLTE effects in the most metal-poor stars may require further investigation.

NLTE analysis is crucially important for lithium studies as well. 
Recent studies have found that NLTE impacts line formation in 1D stellar atmosphere models when lithium abundance is extremely high \citep[e.g.,][]{2018NatAs...2..790Y, 2018ApJ...852L..31L}. Meanwhile, combining NLTE with 3D models has refined the lithium abundance determinations for a massive amount of stars \citep{2021MNRAS.500.2159W}. 
Finally, \cite{2015ApJ...808..148S} and \cite{2025ApJ...983..127S} revealed systematic $\teff$ overestimation in LTE models, which may skew lithium abundance trends. 
By resolving such biases, NLTE abundance analysis may help to clarify the `breakdown' of the Spite plateau at low metallicities and bridge observational data and theoretical models in the quest to resolve the lithium problem.

Given these facts, systematic NLTE analyses to both stellar parameters and elemental abundances of a large sample of metal-poor stars are of great importance. Such efforts have been done by a number of works. For example, \cite{2015ApJ...808..148S} and \cite{2016ApJ...833..225Z} presented systematic NLTE analyses to a large sample of metal-poor stars 

observed by the Lick telescope. Their program stars are down to the metallicity of $\sim -2.4$\,dex. Later, \cite{2022ApJ...931..146A}, \cite{2022ApJ...931..147L} and \cite{2024ApJ...966..174Z} carried out a more massive project that they studied over 400 VMP stars found by the LAMOST survey \citep{2012RAA....12.1197C, 2012RAA....12..723Z, 2022Innov...300224Y} and observed by the Subaru telescope, but this work was done in the LTE assumption. A recent study by \citeauthor{2025ApJ...983..127S} (\citeyear{2025ApJ...983..127S}, hereafter Paper I) presented stellar parameters for over 100 VMP stars, taking advantage of NLTE line formation calculations for \ion{Fe}{1} and \ion{Fe}{2}. Paper I is the first paper in this series, which aims to present a systematic NLTE study of VMP stars with metallicity extended down to $\sim -4.3$\,dex.

In this second paper, we present a homogeneous NLTE analysis of the lithium abundances for 103 VMP stars (with stellar parameters presented in Paper I). 
The paper is organized as follows: 
In Sec.~\ref{section:data}, we present how our data are collected and selected; 
In Sec.~\ref{section:methods}; we present the method used for obtaining the stellar parameters, lithium abundance, NLTE corrections, and evolutionary stages for our sample stars; 
In Sec.~\ref{section:results}, we present the lithium abundance result and how we come to our conclusions. 
Then we show the summary in the final section Sec.~\ref{section:summary}.

\section{DATA} \label{section:data}
\subsection{Observation and Sample Selection} \label{subsection:2-1}
We selected our program stars from a large sample that consists of $\sim 400$ very metal-poor stars \citep{2022ApJ...931..146A, 2022ApJ...931..147L, 2024ApJ...966..174Z}, which were originally selected from the LAMOST survey and then observed by Subaru/HDS during the year of 2014-2017 via a joint program. 
The observations were conducted with a resolution power of R $\sim 36,000$. Observation details can be found in \cite{2022ApJ...931..146A}.

\begin{figure}
\centering
\includegraphics[width=\linewidth]{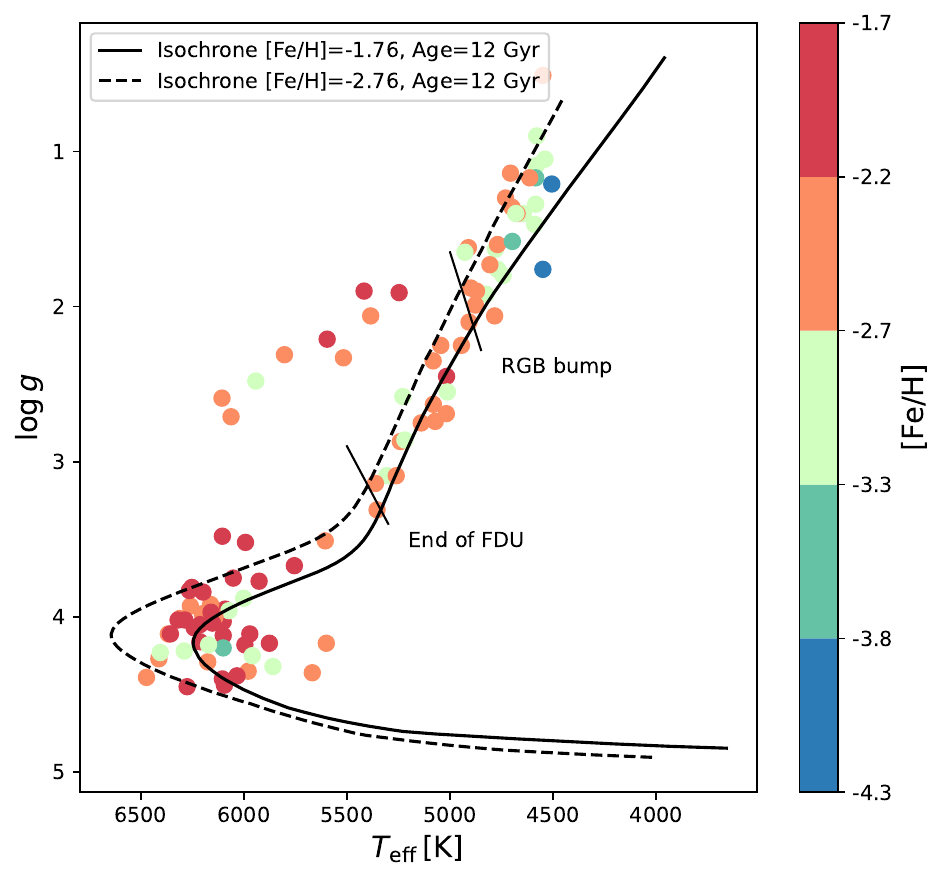}
    \caption{Distribution of our sample stars in the H-R diagram, color-coded by [Fe/H]. The solid and dashed lines represent 12 Gyr $\mathrm{Y^2}$ isochrones with [Fe/H] = $-1.76$ and $-2.76$, respectively. The approximate location of the first dredge-up termination (FDU) and the RGB bump are indicated following the way of \cite{2022A&A...661A.153M}.}
\label{fig:1}
\end{figure}

For a reliable systematic NLTE analysis to VMP stars and a comprehensive investigation to the chemical evolution of the early Milky Way, we selected our program stars by the following criterion: 
1) the signal-to-noise ratio (SNR) of their spectra are over 50 for stars with [Fe/H] $\ge -4.0$ at the range of $4,500$\,\AA; 
2) they must cover the evolutionary stages from the unevolved main sequence phase to the highly evolved phase of horizontal branch (see Paper I); 
3) All stars with [Fe/H] $< -4.0$ are included regardless of their spectral SNR. This procedure yields a sample of 103 stars with metallicity down to $-4.3$\,dex. 
We show our sample stars on a H-R diagram in Fig.~\ref{fig:1}.

\subsection{Stellar Parameters} \label{subsection:2-2}
Stellar atmospheric parameters were adopted from Paper I. These parameters were established with a two‐step approach.
The final uncertainties are typically 50–100\,K in $T_{\rm eff}$, 0.2\,dex in $\log g$, 0.1\,dex in [Fe/H], and 0.2\,km\,s$^{-1}$ in $\xi_{t}$ (see Paper I for details), providing a solid foundation for our subsequent elemental abundance determinations.

\section{Methods} \label{section:methods}

\subsection{Measurement of Li Abundance} \label{subsection:3-1}
We use the spectrum synthesize method to derive the lithium abundances. 
Spectral synthesis analysis used an IDL-based software package {\it SIU} \citep{1999PhDT.......216R} and grids of MARCS atmospheric models \citep{2008A&A...486..951G}. 
For most of our program stars, lithium abundances were measured using the resonance \ion{Li}{1} doublet at $6707.76$ and $6707.91$\,\AA. For the four Li-rich stars in our sample, the subordinate line at $6103.6$\,\AA\ is strong enough to derive reliable lithium abundances. In these four stars, both the resonance line and subordinate line were used.
The data of the atomic line are listed in Table~\ref{tab:1}. 

\begin{table}[t]
    \centering
    \caption{Li atomic data}
    \begin{tabular}{ccrl}
    \hline\hline
        {Wavelength} & {Transition} &  {$\log{gf}$\tablenotemark{a,b}} & {$\log C_{6}$\tablenotemark{c}}  \\ \hline
       $6707.76$ &  $1s^22s-1s^22p$ & $-0.002$ & $-31.255$ \\
        $6707.91$ &  $1s^22s-1s^22p$ & $-0.299$ & $-31.255$ \\
        $6103.54$ &  $1s^23d-1s^22p$ & $0.101$ & $-30.317$ \\
        \hline
    \end{tabular}
    \label{tab:1}
    \tablenotetext{}{References: a.\citet{2007A&A...465..587S} b.\citet{2018NatAs...2..790Y} c.\citet{1998MNRAS.296.1057B}}
\end{table}

The lithium abundance is expressed as A(Li): $A(Li)=\log(N_{\rm Li}/N_{\rm H})+12$, where $N_{\rm Li}$ and $N_{\rm H}$ represent the number densities of lithium and hydrogen in stars, respectively. 
To fit the observed spectral line profiles, spectral broadening caused by rotation, macroscopic turbulence, and instrumental broadening has been taken into account as a Gaussian broadening profile. 

\subsection{NLTE Calculation} \label{subsection:3-2} 
\begin{figure*}[ht]
  \centering
\includegraphics[width=\linewidth]{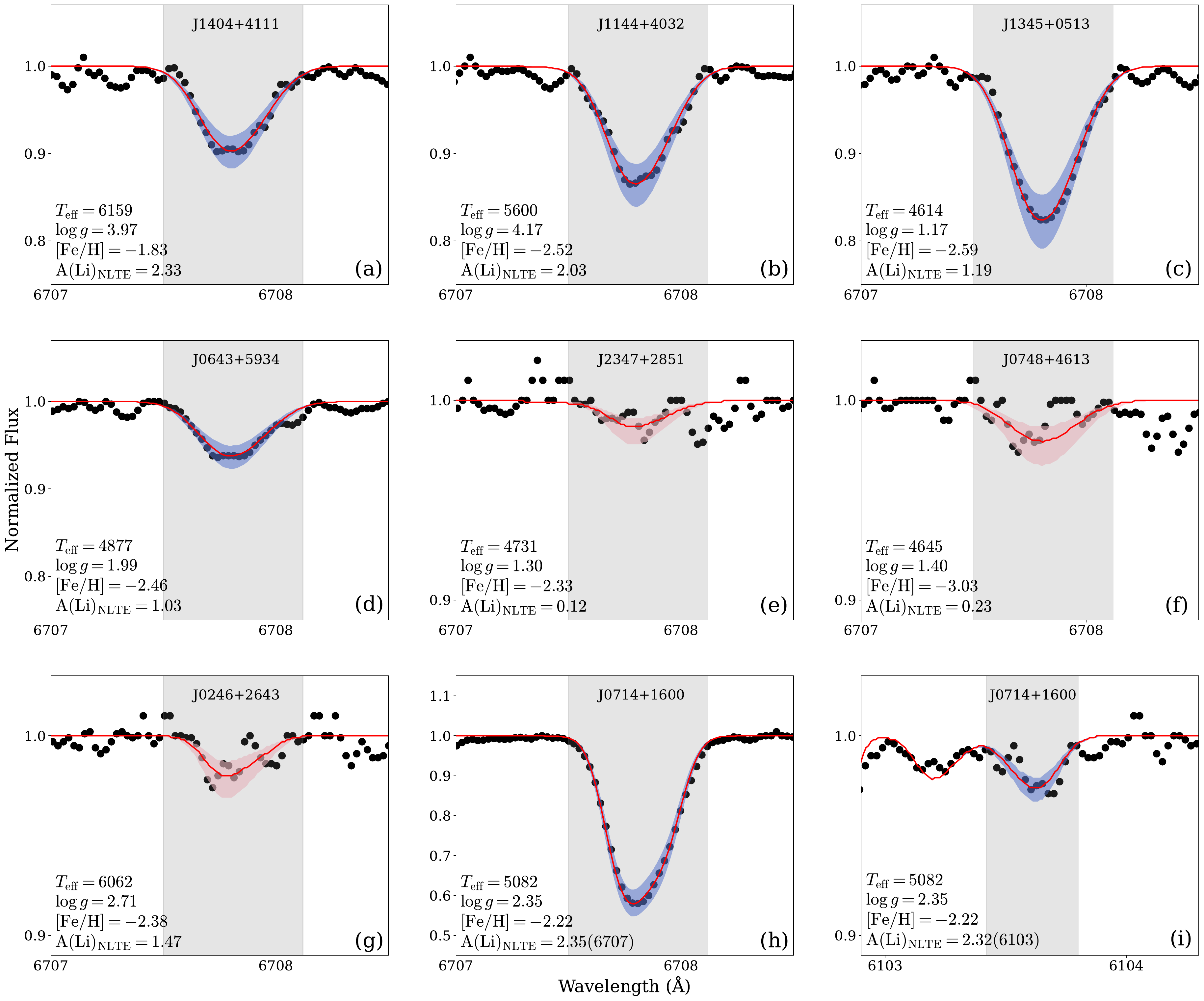}
   \caption{The examples of line profile fitting in this work. In all the panels, the black dots are the observed spectra and the red lines are the theoretical spectra. Panels (a) and (b) show two fitting examples of unevolved stars. Panels (c) and (d) show two fitting examples of LRGB stars. Panels (e) and (f) show `upper limit' fittings of two highly evolved RGB stars, and panel (g) shows `upper limit' fitting to a high temperature HB star. Panels (h) and (i) show fittings to resonance line and subordinate line of a Li-rich star, respectively. The blue areas represent synthesis spectra with $\rm{A(Li)_{{NLTE}}}\pm0.1\,dex$ in panels (a),(b), (c), (d), (h) and (i), while the red areas represent synthesis spectra with $\rm{A(Li)_{{NLTE}}}\pm0.2\,dex$ in panels (e),(f), and (g).}
   \label{fig:2}
\end{figure*}

NLTE effects were taken into account in our determination of lithium abundances.
We used the lithium atomic model developed by \cite{2007A&A...465..587S}.
To solve the coupled radiative transfer and statistical equilibrium equations, we employ a revised version of the {\it DETAIL} program based on the accelerated lambda iteration method 
\citep{1991A&A...245..171R,1992A&A...262..209R}, as updated by \citet{2011A&A...528A..87M}.
The resulting departure coefficients were then utilized by {\it SIU} to compute NLTE synthetic line profiles. In Fig.~\ref{fig:2}, we present several examples of the line profile fitting results at the resonance \ion{Li}{1} lines of both LTE and NLTE analysis. 
The NLTE effects for Li abundance are defined and calculated by A(Li)$_{\rm NLTE}-$A(Li)$_{\rm LTE}$.
It is important to note that, in stars with extremely high lithium abundances, the \ion{Li}{1} lines are strong, and LTE-based line profile fitting fails to reproduce the observed profiles, whereas NLTE calculations provide a much better match.

\subsection{Determination of Evolution Phase} \label{subsection:3-3}
To assess the evolutionary phase of our sample stars, we compare their observed effective temperatures and surface gravities with theoretical isochrones computed at an age of 12\,Gyr. 
The isochrone models adopted in this work are drawn from the grid presented by \citet{2004Ap&SS.291..261Y}, which assumes a fixed [\(\alpha\)/Fe] enhancement of 0.4~dex. 
However, because the available isochrone models are produced only at a few discrete metallicities, an isochrone corresponding to a specific iron abundance cannot be obtained directly.

To overcome this, we generate the desired 12\,Gyr isochrone by linearly interpolating between the two bracketing models. 
In our approach, the measured [Fe/H] is first converted to a corresponding metallicity; then, when the target metallicity does not exactly match one of the discrete model values, an interpolation is performed between the model that is just higher and that just lower than the target. 
This procedure yields a synthetic isochrone that reflects the target metal content.

We apply this interpolation method to our sample, which is divided into four [Fe/H] bins. 
These bins are defined as follows: stars with [Fe/H] $>-2.5$, stars with [Fe/H] between $-2.5$ and $-3.0$, stars with [Fe/H] between $-3.0$ and $-3.5$, and stars with [Fe/H] $<-3.5$. 
For each bin, a representative [Fe/H] value is adopted (namely $-2.25$, $-2.75$, $-3.25$, and $-3.75$, respectively) to perform the interpolation.

\section{Results \& Discussion} \label{section:results}

\subsection{Lithium Abundance and Uncertainty Estimation} \label{subsection:4-1}
We obtained the LTE and NLTE lithium abundances for 77 stars in our sample. For 23 of the remaining stars, their lithium lines are not very clear but marginally recognizable, we performed a fitting using slightly stronger line profiles and adopted the derived abundances as upper limits for A(Li), and most of these stars are highly evolved stars. For the last three stars, we cannot detect any lithium features in their spectra, and we do not present lithium abundance or upper limit for these three stars.
The results of Li abundance obtained from both LTE and NLTE analysis are listed in Table~\ref{tab:2} (and Table~\ref{tab:a1} in the Appendix) together with the stellar parameters. 
Most of the lithium abundance and all the upper limits shown in Tables \ref{tab:2} and \ref{tab:a1} are derived from the resonance \ion{Li}{1} line, as we previously mentioned, while for the four Li-rich stars, their abundances are an averaged value of the resonance \ion{Li}{1} line and the subordinate line. The differences between these two lines are from 0.03 to 0.20\,dex with an average of 0.12\,dex for these four stars.

\begin{figure}[h]
\includegraphics[width=\linewidth]{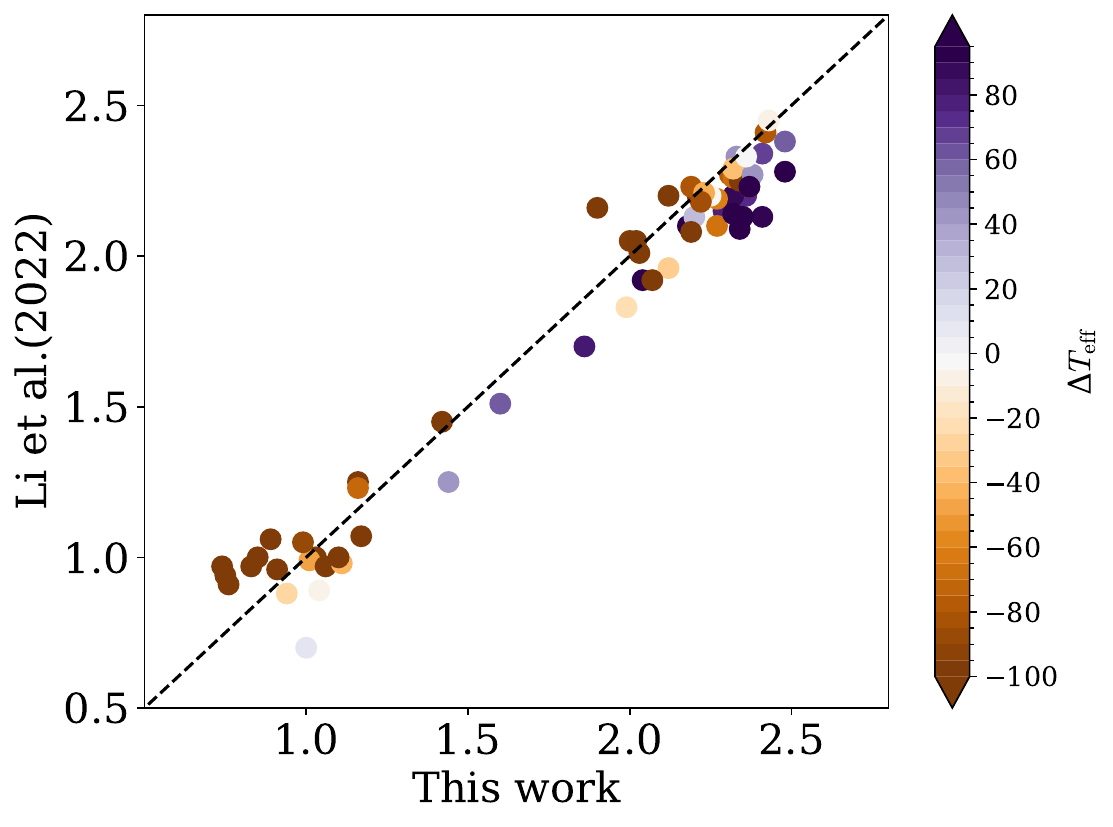}
   \caption{Comparison of the LTE Li abundances with our previous work \citep{2022ApJ...931..147L}. The difference in effective temperatures (This work $-$ literature) are indicated with a color-bar on the right. Li-rich stars are not included in this comparison, as the LTE abundances in super Li-rich stars are not reliable.}
   \label{fig:3}
\end{figure}

Our sample overlaps with that of \cite{2022ApJ...931..147L}, which presented the lithium abundance in LTE. In Fig.~\ref{fig:3}, we show a comparison of the LTE lithium abundances with the previous work. Li-rich stars are not included in this comparison, as the LTE abundances in super Li-rich stars are not reliable. For most stars, they show a strong connection with the difference in effective temperatures, which is expected since the abundance of lithium is sensitive to $\teff$. The mean difference is 0.05\,dex with a scatter of 0.12\,dex.

\begin{figure*}[ht]
  \centering
\includegraphics[width=\linewidth]{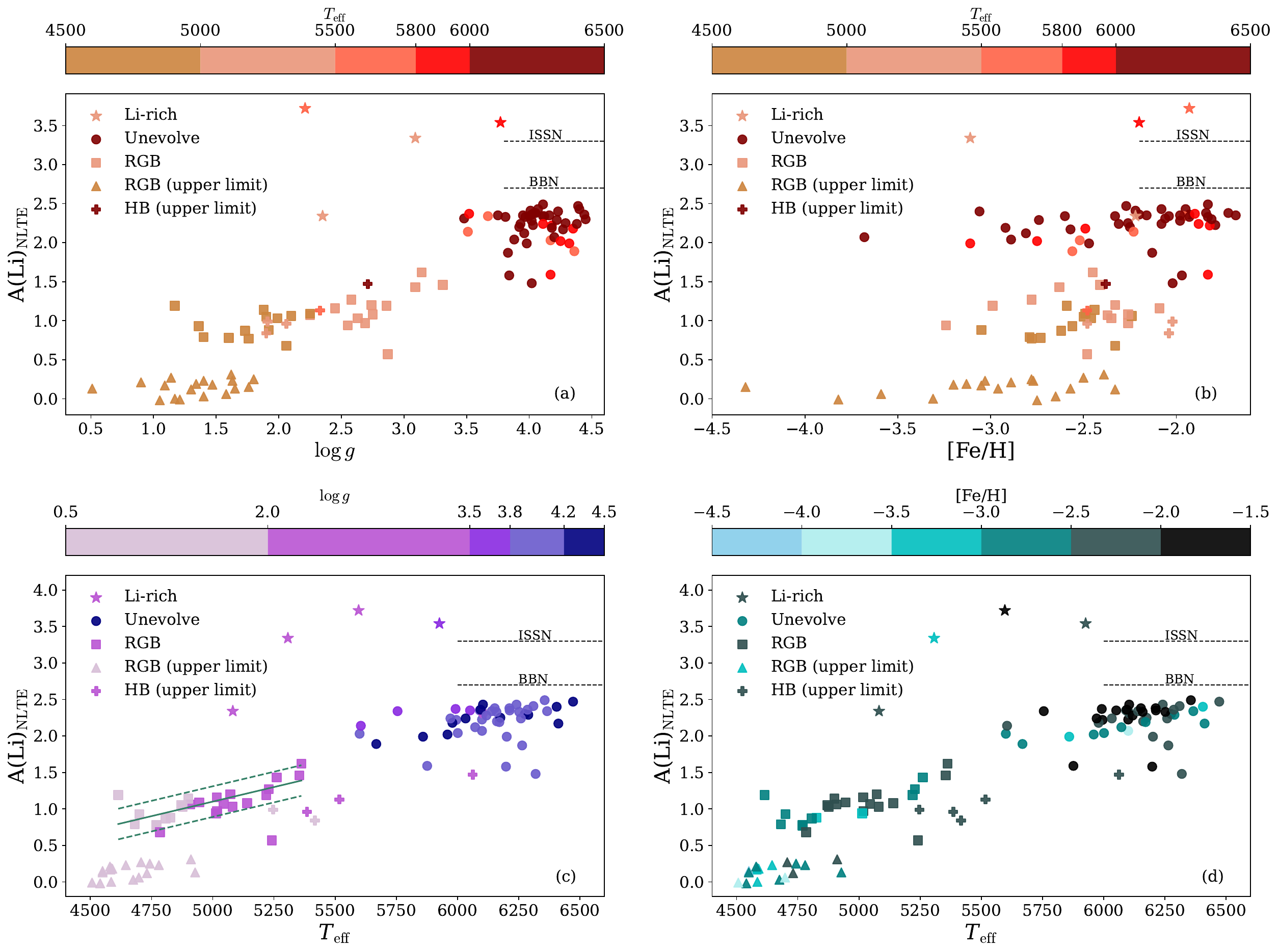}
   \caption{The NLTE Li abundances as functions of stellar parameters. Panels (a) and (b) show the NLTE Li abundance as functions of surface gravity and metallicity, with effective temperature as a color bar. Panels (c) and (d) show the NLTE Li abundance as a function effective temperature, with surface gravity and metallicity as color bars, respectively. The horizontal dashed lines with BBN represents with the Li abundance predicted by the standard BBN theory, while the other horizontal dashed lines with ISSN represents the Li abundance of initial solar-system nebula. The four Li-rich stars are indicated as pentagons. The carbon enhanced metal-poor (CEMP) stars are indicated with a gray circle surrounding specific symbols. The lithium gap can be seen in panel a of this figure.}
   \label{fig:4}
\end{figure*}

In Fig.~\ref{fig:4}, we show the NLTE Li abundance as a function of the effective temperature of our program stars. 
Given the wide range of metallicity in our sample stars, the overall trend is consistent with our previous work \citep{2016ApJ...833..225Z} and many other works \citep[e.g.,][]{2022ApJ...937...52B}. 
Four outliers show anomaly high Li abundance compared with other stars at their evolutionary stages, which are identified as Li-rich stars (detailed discussions are presented in Sec.~\ref{subsection:4-4}). 
Except for the Li-rich stars, all the other stars show a Li abundance that is significantly lower than the predicted value from BBN. 
The Spite plateau can be clearly seen from Fig.~\ref{fig:4} panel (b), and we present these details in Sec.~\ref{subsection:4-3}.

\begin{table*}[t]
    \centering
    \caption{Li Abundance and Stellar Parameters}
    \begin{tabular}{lccrcccc}
    \hline\hline
        {Star} & {$T_{\rm{eff}}$(K)} &  {$\log{g}$(dex)} & {[Fe/H]} &{$\xi_{\rm t}(km\,s^{-1})$}& {A(Li)$_{\rm{LTE}}$} &{A(Li)$_{\rm{NLTE}}$} & {phase}  \\ \hline
J$0055+1857$ & 5018 & 2.69 & $-$2.26 & 1.40 & 0.94 & 0.97 & 2  \\ 
J$0119+2425$ & 6412 & 4.27 & $-$2.57 & 1.64 & 2.18 & 2.17 & 1  \\ 
J$0131+4800$ & 6100 & 4.03 & $-$1.79 & 1.35 & 2.21 & 2.23 & 1  \\ 
J$0232+0545$ & 6091 & 3.95 & $-$2.16 & 1.11 & 2.34 & 2.35 & 1  \\ 
J$0244+0828$ & 6472 & 4.39 & $-$2.27 & 1.60 & 2.48 & 2.47 & 1  \\ 
J$0246+2643$ & 6062 & 2.71 & $-$2.38 & 3.30 & 1.44 & 1.47 & 4  \\ 
J$0326+0202$ & 4827 & 1.92 & $-$3.05 & 1.52 & 0.85 & 0.88 & 3  \\ 
J$0352+0514$ & 5859 & 4.32 & $-$3.11 & 1.20 & 1.99 & 1.99 & 1  \\ 
J$0423+0538$ & 6100 & 4.12 & $-$2.04 & 1.31 & 2.33 & 2.34 & 1 \\ 

...  & ... & ... & ... & ... & ...  & ...   \\
        \hline
    \end{tabular}
    \label{tab:2}
    \tablenotetext{}{Note: 1.unevolved stars; 2. RGB stars; 3. highly evolve RGB stars; 4. horizontal branch.}
\end{table*}


\begin{figure*}
  \centering
\includegraphics[width=0.8\linewidth]{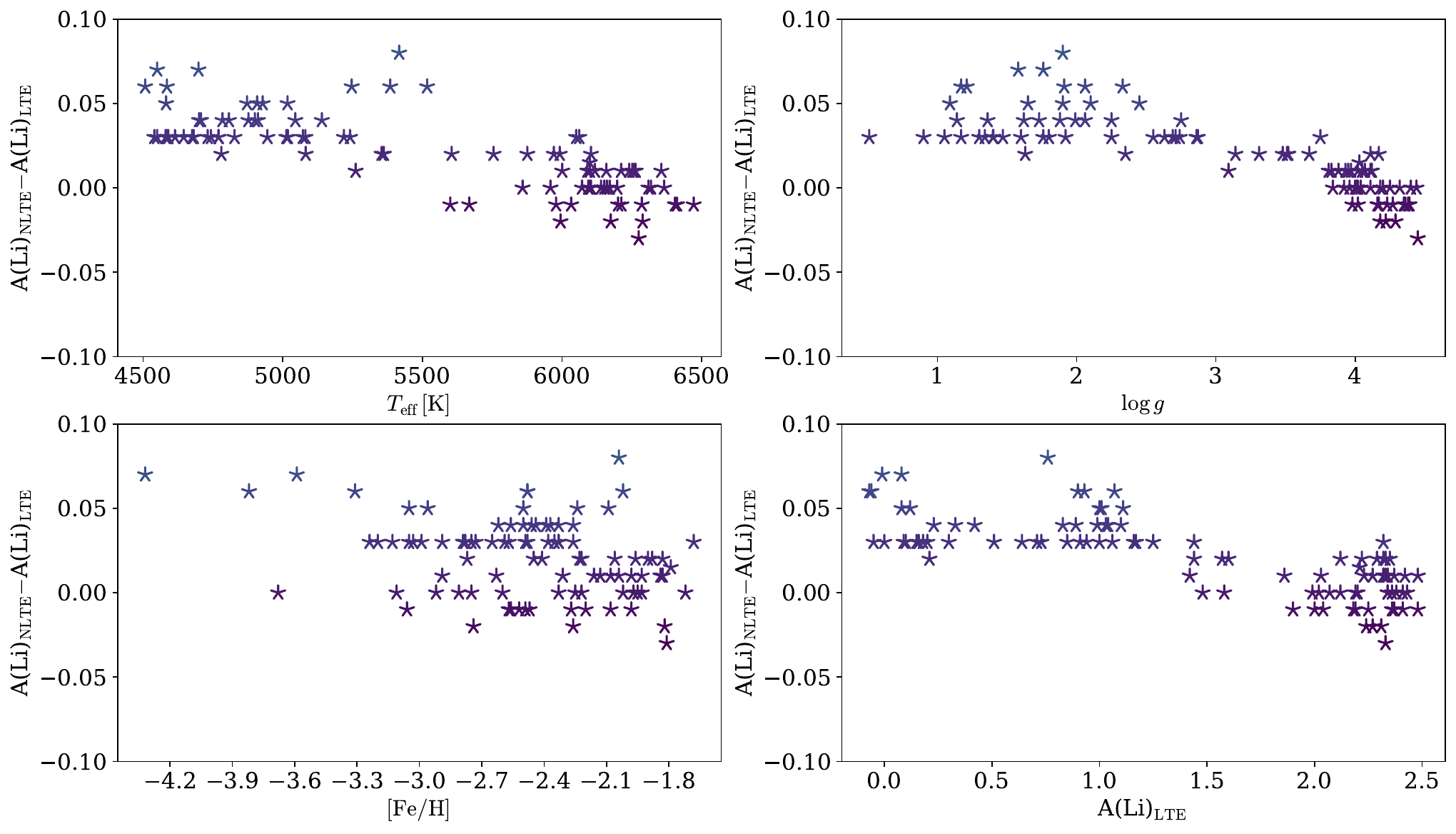}
   \caption{$\rm{A(Li)_{{NLTE}}}-\rm{A(Li)_{{LTE}}}$ of our work versus stellar parameters and A(Li)$_{\rm{LTE}}$. The result of $\rm{A(Li)_{{NLTE}}}-\rm{A(Li)_{{LTE}}}$ for the three most Li-rich stars: J$0554+5235$, J$0626+6032$, and J$0705+2552$  are -0.32, -0.18, -0.60\,dex, respectively. These Li-rich stars are not shown in Fig.~\ref{fig:5} to avoid clutter and to provide a clearer view of the corrections for the majority of the stars.}
    \label{fig:5}
\end{figure*}

In Fig.~\ref{fig:5}, we show the NLTE effects of the Li resonance lines at $6708$\,\AA\ in our sample stars as functions of $T_{\rm eff}$, $\log g$, [Fe/H], and A(Li)$_{\rm LTE}$. 
For most of our program stars, the NLTE effects are not very large, and the absolute values of the difference are at a level of $\sim 0.05$\,dex. For the four Li-rich stars, this effect can reach $\sim 0.60$\,dex, which is consistent with what is reported by previous work \cite[e.g.,][]{2021NatAs...5...86Y}. To avoid clutter and to provide a clearer view of for the majority of the stars, the NLTE effects of the three most Li-rich stars are not shown in Fig.~\ref{fig:5} but presented in the caption of the figure.

The uncertainty of the Li abundance was mainly caused by the uncertainties of stellar parameters (namely $\teff$, $\log {g}$, [Fe/H] and $\vt$). 
Thus, we tested three stars from the VMP sample by varying the stellar parameters within their typical uncertainties, which are $\Delta T_{\rm{eff}} = \pm100$\,K, $\Delta \log {g} = \pm0.10$\,dex, and $\Delta$[Fe/H] $= \pm0.1$0\,dex, $\Delta \vt = \pm0.2$\,\kms, respectively.
The results are shown in Table~\ref{tab:3}.
The Li abundance is more sensitive to the $T_{ \rm{eff}}$ than that of other stellar parameters. 
Our final uncertainty of the lithium abundance is estimated from the weighted average of the errors due to the uncertainties in the three parameters.

\begin{table} \label{tab:3}
\centering
\caption{Li abundance errors due to the uncertainties in the stellar parameters}
\begin{tabular}{lcrrr}
\hline \hline
    & Input error & J2217 & J1404 & J0554  \\ \hline
                                & {$\teff$\,(K)} & 4507 & 6159 & 5596 \\
\multirow{2}{*}{$\Delta$A(Li)}  & $+100$ & $+0.11$& $+0.07$ &$+0.10$\\
                                & $-100$ & $-0.08$ &$-0.07$ & $-0.10$\\ \hline
                                & {$\log$g\,(dex)} & 1.21 & 3.97 & 2.21 \\
\multirow{2}{*}{$\Delta$A(Li)}  & $+0.10$ & $+0.01$ & $0.00$ & $0.00$ \\
                                & $-0.10$ & $-0.01$ & $0.00$ & $0.00$\\\hline

                                & [Fe/H]\,(dex) & $-3.82$ & $ -1.83$ & $-1.93$ \\
\multirow{2}{*}{$\Delta$A(Li)}  & $+0.10$ & $0.00$ & $0.00$ & $-0.01$ \\
                                & $-0.10$ & $0.00$ & $0.00$ & $+0.01$ \\\hline
                                & $\xi_t$\,(\kms) & 2.40 & 1.35 & 2.30 \\
\multirow{2}{*}{$\Delta$A(Li)}  & $+0.20$ & $+0.01$ & $0.00$ & $0.00$ \\
                                & $-0.20$ & $0.00$ & $0.00$ & $0.00$ \\
                       \hline
\end{tabular}
\end{table}

\subsection{Evolutionary Phases} \label{subsection:4-2}

\begin{figure*}
\centering
\includegraphics[width=\linewidth]{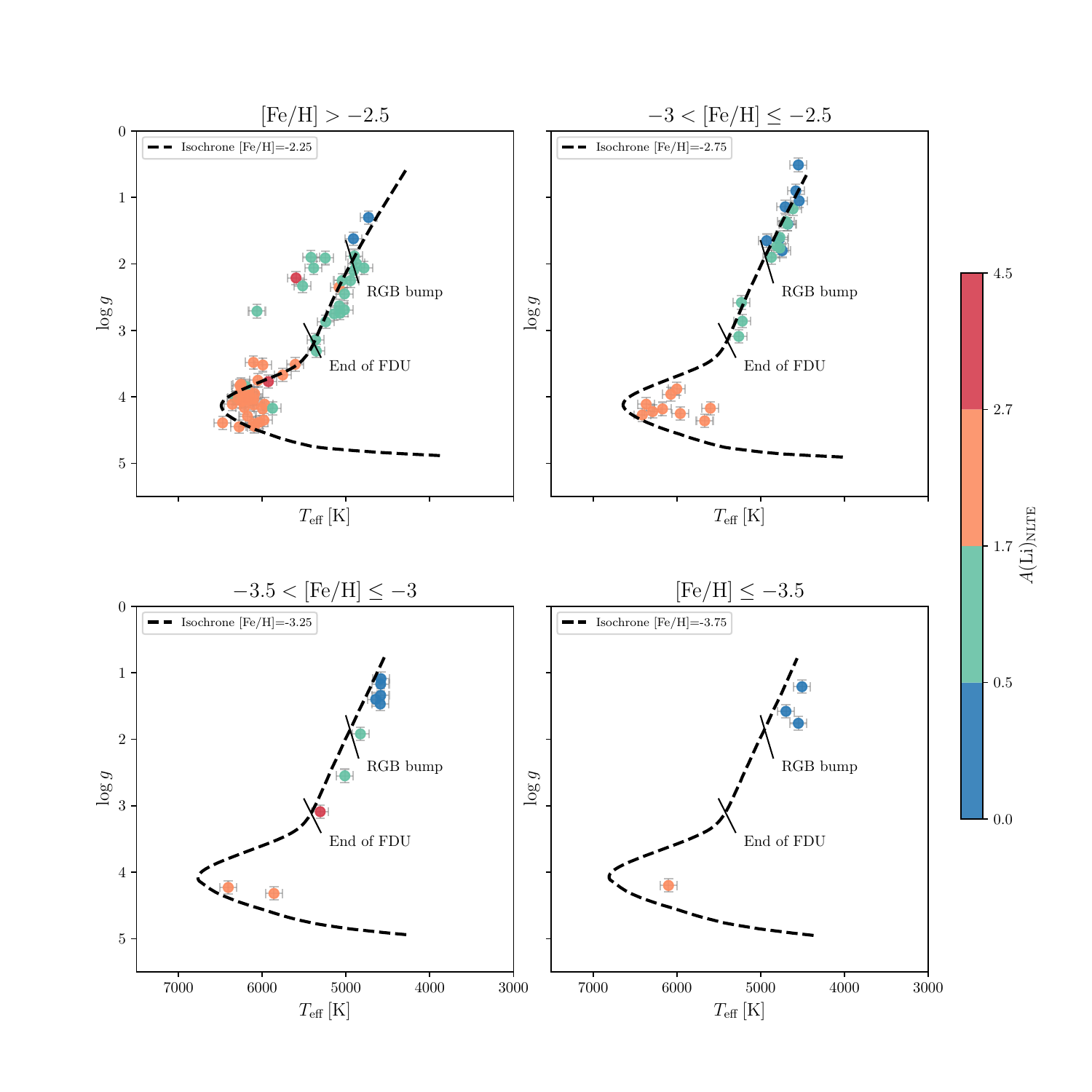}
\caption{12\,Gyr $\mathrm{Y^2}$ isochrones for stars grouped by metallicity, shown in four panels with different [Fe/H] intervals: [Fe/H] $> -2.5$ (isochrone interpolated at [Fe/H] $=-2.25$), $-2.5 \geq \mathrm{[Fe/H]} > -3.0$ (isochrone at $-2.75$), $-3.0 \geq \mathrm{[Fe/H]} > -3.5$ (isochrone at $-3.25$), and [Fe/H] $< -3.5$ (isochrone at $-3.75$). Dashed lines represent the interpolated theoretical isochrones, and filled colored circles denote the observed stellar parameters, with the color bar indicating A(Li)$_\mathrm{NLTE}$. As in Fig.\ref{fig:1}, each panel marks the approximate positions of the RGB bump and the end of FDU.
}

\label{fig:6}
\end{figure*}

Fig.~\ref{fig:6} presents the isochrones of our program stars and their lithium abundances in a 2$\times$2 layout. In each panel, the dashed line shows the interpolated 12\,Gyr isochrone appropriate for the designated [Fe/H] bin, while the filled black circles denote the observed stellar parameters. The effective temperature and surface gravity axes are inverted following standard astronomical conventions. The overall agreement between the observed positions and the interpolated isochrones confirms the robustness of our spectroscopic measurements and provides a reliable means to determine the evolutionary phases of stars over a wide metallicity range. For a better classification, we also marked the rough positions of the end of the first dredge-up (FDU) and RGB-bump in each panel, respectively \citep{2022A&A...661A.153M}. 
Based on the positions of our program stars, they are classified as four groups: the unevolved stars - stars that have not reached FDU, lower RGB (LRGB) stars - stars between FDU and RGB-bump, highly evolved stars - stars that passed the RGB-bump, and the horizontal branch (HB) stars - stars to the left of the dashed isochrone lines. 

\subsection{The Lithium Plateaus and the Lithium Gap} \label{subsection:4-3}

\subsubsection{The Spite Plateau and Its Meltdown in Unevolved Stars} \label{subsubsection:4-3-1}

The lithium abundance in warm, metal-poor Population II stars is commonly regarded as a proxy for the primordial value \citep{1982Natur.297..483S}.
For our data as seen in Fig.~\ref{fig:4}, the unevolved warm metal-poor halo stars with effective temperature between $6,500$\,K and $6,000$\,K and [Fe/H] between $-2.5$\,dex and $-1.7$\,dex show a Li plateau (Spite plateau) at A(Li) $= 2.32$\,dex (hereafter referred to as plateau stars).
This value is consistent with many previous studies \citep[e.g.,][]{1987A&A...172L..17R, 1997MNRAS.285..847B, 1999A&AS..137...93G, 2000ApJ...530L..57R, 2007A&A...462..851B, 2018A&A...612A..65B, 2003ChJAA...3..453Z, 2004ApJ...615L..33M, 2007A&A...465..587S, 2009ApJ...698.1803A, 2009A&A...493..601H, 2010A&A...519L...3M, 2010A&A...522A..26S, 2012PASP..124..164S, 2016ApJ...833..225Z, 2020MNRAS.496.2902M, 2021MNRAS.506.1438K}.

\begin{figure*}
\centering
\includegraphics[width=0.8\linewidth]{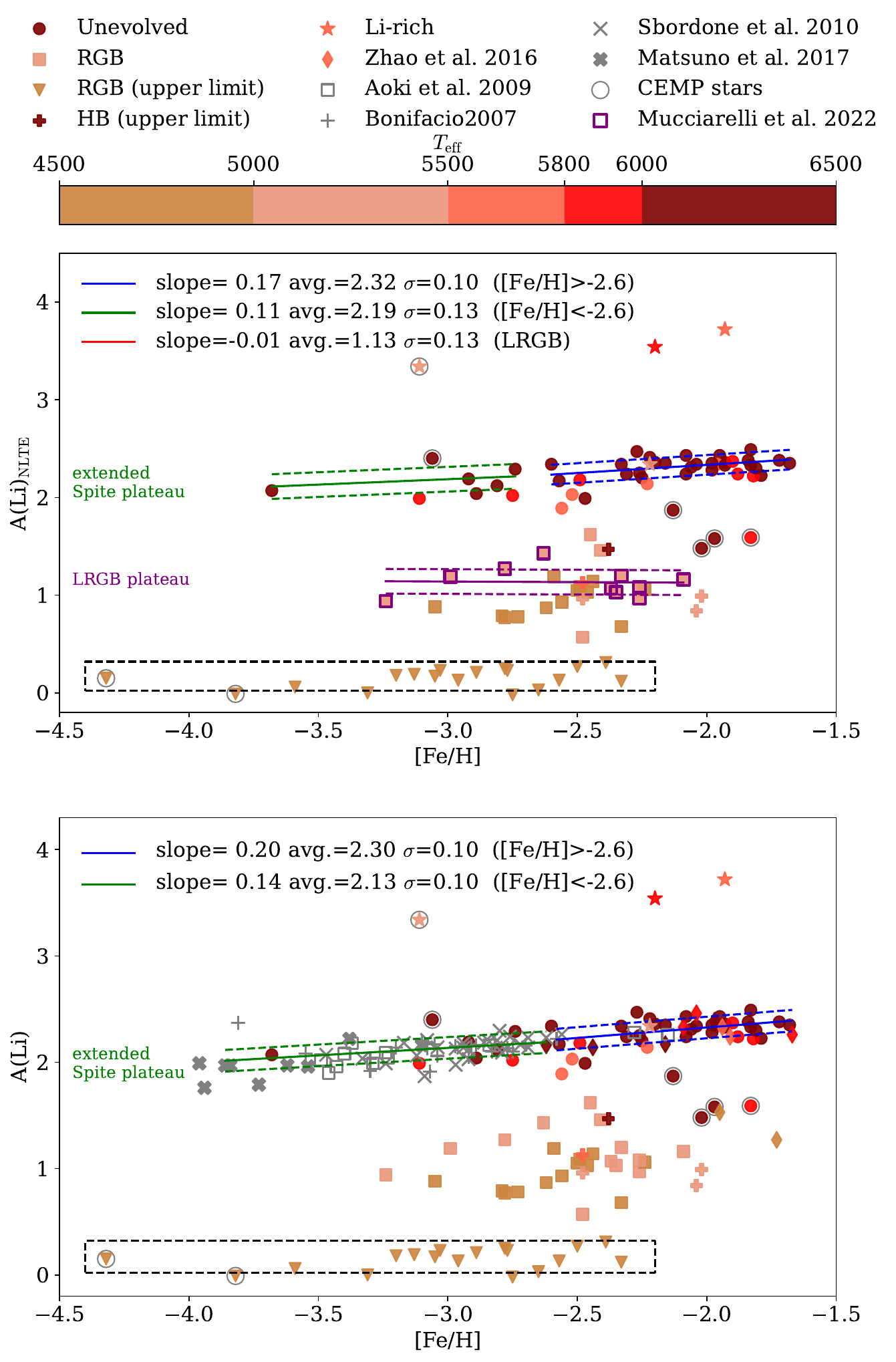}
\caption{The NLTE lithium abundance and fittings to lithium plateaus. The upper panel shows stars in our sample, while the lower panel shows stars in our sample and from literature \citep{2007A&A...462..851B,2009ApJ...698.1803A,2010A&A...522A..26S,2017AJ....154...52M, 2016ApJ...833..225Z}. Symbols are the same with Fig.~\ref{fig:4}. Squares outlined in thick purple indicate the subsample selected following \citet{2022A&A...661A.153M}. CEMP stars are indicated with a gray circle surrounding the specific symbols.  The lithium gap can be seen in all panels of this figure and panel a of Fig.~\ref{fig:4}.}

\label{fig:7}
\end{figure*}

However, there is an argument that if the Spite plateau is strictly flat. 
For a better investigation of the Spite plateau, in addition to Fig.~\ref{fig:4}, we show the Li abundances (including ours and those in the literature) as functions of [Fe/H] and the linear fitting to the trend in Fig.~\ref{fig:7}. 
In the upper panel of Fig.~\ref{fig:7}, we present a linear fit to the unevolved plateau stars ($\log g \ge 4.0$). The best linear fit is A(Li)$= (0.17 \pm 0.07) \times$[Fe/H]+2.67. The standard deviation of the lithium abundances in these unevolved plateau stars is $0.10$\,dex. Under both LTE and NLTE assumptions, the trends of A(Li) with [Fe/H] remains the same, with only minor differences in the fitted slopes given that NLTE corrections for our sample stars as a function of [Fe/H] show no obvious trend as seen in Fig.~\ref{fig:5}. Similarly, 3D effects are not expected to introduce a strong systematic trend with metallicity.

It has been reported that the Spite plateau breaks for stars with metallicity lower than $\sim -2.5$\,dex \citep[e.g.,][]{2009ApJ...698.1803A, 2010A&A...522A..26S}. 
The deviations from the Spite plateau can be interpreted as hints at mechanisms of lithium depletion acting during the lifetime of Population II stars \citep{2023lau..book.....M}. 
In our program stars, there is indeed a decline and larger scatter of lithium abundance towards the metal-poor end, however, as seen from the upper panel of Fig.~\ref{fig:7}, lithium abundance in stars with effective temperatures higher than $6,000$\,K (warm) remains in a thin plateau with very narrow variations. 
For our warm unevolved sample stars, there is a small gap in metallicity of $\sim -2.6$. Thus, we perform two linear fittings to warm stars with a metallicity higher and lower than $-2.6$\,dex, respectively. 
Interestingly, they show a very similar slope (0.17 vs. 0.11) and A(Li) scatter (0.10\,dex vs. 0.13\,dex), respectively (see the blue and green lines in the upper panel of Fig.~\ref{fig:7}). 
To rule out the sample effect, we extended the sample by including more stars from the literature \citep{2007A&A...462..851B,2009ApJ...698.1803A,2010A&A...522A..26S,2017AJ....154...52M,2016ApJ...833..225Z}, and performed the same fitting procedure, as shown in the lower panel of Fig.~\ref{fig:7}.
We found a similar result: the classic Spite plateau shows a slope of 0.20 with a A(Li) scatter of 0.10\,dex, and the `meltdown' stars show a slope of 0.14 with a A(Li) scatter of 0.10\,dex. Given this result, one can estimate that the average lithium abundances at [Fe/H] $= -3.0$ and $= -3.5$ will be lower than those at [Fe/H] $= -2.0$ by $\sim 0.17$\,dex and $\sim 0.24$\,dex, which is consistent with the value reported in previous work \cite[e.g.][]{1999ApJ...523..654R, 2009ApJ...698.1803A, 2017AJ....154...52M}.

The overall trend of metallicity-extended Spite plateau decreases towards the metal-poor direction. 
This result suggests that the classic Spite plateau is not strictly flat; most significantly, it is likely to extend to lower metallicities as some previous work suggested \citep[e.g.,][]{2017AJ....154...52M}. 
However, the behavior of lithium abundances below [Fe/H] $< -3.0$ remains debated.
The so-called `meltdown' of the Spite plateau has been interpreted either as an increased scatter in lithium abundances or as a systematic decline toward lower metallicity.
It should be mentioned that in our sample, only one unevolved star reaches [Fe/H] $\sim -3.6$, which prevents a firm conclusion regarding the presence or absence of increased scatter in this region.
Recent studies of extremely metal-poor stars \citep[e.g.,][]{2019ApJ...874L..21A, 2021A&A...654A.170M} have reported diverse lithium behaviors at the lowest metallicities. 
Further investigations with larger samples are required to clarify whether a true breakdown of the plateau occurs.

The slightly positive slope of the extended Spite Plateau may indicate that, in addition to the primordial component, a small amount of Galactic lithium production has already taken place at early times.
Nova has been suggested to be one of the main contributors to the Galactic lithium enrichment \citep[e.g.,][]{1975A&A....42...55A, 1978ApJ...222..600S, 2024ApJ...962..191S, 1996ApJ...465L..27H, 1998ApJ...494..680J, 2014MNRAS.442.2058D, 2015Natur.518..381T, 2015ApJ...808L..14I, 2016MNRAS.463L.117M, 2021ApJ...916...44A, 2022MNRAS.509.3258M, 2018MNRAS.481.2261S, 2022ApJ...933L..30K, 2024ApJ...971....4G, 2025A&A...699A.171M}.
While nova enrichment is expected to become important at higher metallicities \citep{2021A&A...653A..72R, 2024A&A...691A.142B}, current models remain relatively unconstrained in the metal-poor regime.
Given the substantial lithium yield of individual events, early novae could still provide a non-negligible contribution to lithium enhancement in VMP stars considered in this work.
Additionally, cosmic-ray spallation operates from early epochs and behaves predominantly as a metallicity-dependent process \citep[e.g.,][]{1998ApJ...499..735L, 2019MNRAS.489.3539G}. 
While such behavior is qualitatively consistent with an increase of lithium abundance with increasing [Fe/H], chemical evolution models \citep[e.g.,][]{2012A&A...542A..67P} suggest that its contribution at very low metallicity is limited, and may not be sufficient alone to account for the observed trend in our sample. 
A joint contribution from both early Galactic enrichment mechanisms may be required.

Meanwhile, we note that such a slope could also be interpreted as a signature of metallicity-dependent stellar depletion. While the Spite plateau is generally considered to be depleted from the primordial value (A(Li) $\simeq 2.7$) due to the processes such as diffusion and macro-turbulence \citep[e.g., see][for a recent review]{Charbonnel2026}, a scenario that the depletion efficiency decreases with increasing metallicity could also naturally reproduce this positive slope.

The observed lithium trends in VMP stars likely result from a complex interplay between these internal stellar processes and early Galactic enrichment. Further observational constraints and detailed chemical evolution modeling will be essential to clarify the origin of this trend

\subsubsection{The `Plateau' in Lower RGB Stars} \label{subsubsection:4-3-2}
With stars evolving to the RGB phase, the Li abundance in our sample exhibits additional features.
The apparent dependence on surface gravity seen in Fig.~\ref{fig:4} largely reflects the underlying correlation with effective temperature, which is the dominant parameter in this region.
More than half of the RGB stars in our sample have surface gravities between $\sim 2.0$\,dex and $\sim 3.0$\,dex and effective temperatures from $\sim 4800$\,K to $\sim 5400$\,K. 
In Fig.~\ref{fig:6}, we identify these stars lying between the end of FDU and RGB-bump, which is classified as LRGB. 
Previous work has reported the clue or evidence for the existence of a LRGB Li plateau, For example, \cite{2009A&A...503..545L} showed that the Li abundances of LRGB stars in the metal-poor globular cluster NGC 6397 are around $1.10$\,dex. 
\cite{2012MNRAS.419.2195M, 2022A&A...661A.153M} found a thin plateau among LRGB stars with an average A(Li) = $1.09 \pm 0.01$ dex and a dispersion of $\sigma = 0.07$ dex. These LRGB stars spans a relatively narrow effective temperature range of $\sim 5,000-5,350$ K and a large range of metallicity from from $-1.7$ down to $-7.0$\,dex. They also identified a small fraction of such stars with A(Li) $< 0.7$\,dex.
\cite{2022ApJ...931..147L} reported that the Li abundance of giants remains more or less constant after the first dredge-up to the RGB-bump.

In our program stars, we found a similar plateau in LRGB stars. 
Considering \citet{2022A&A...661A.153M}'s effective temperature range of $4,800$\,K to $5,300$\,K as they shown in their Figure 2, the stars in our sample (marked with a thick purple open square in Fig.~\ref{fig:7}, top panel) show a mean lithium abundance of 1.13$\pm 0.13$\,dex,  with almost a zero slope in A(Li) -- [Fe/H] plane. Additionally, a fitting to the lithium abundance as a function of effective temperature shows a non-zero slope, which is $0.12 \pm 0.03$, slightly larger than that reported by \citep{2022A&A...661A.153M}. Extending the fitting range to a larger effective temperature range (e.g., from $4,800$\,K to $5,400$\,K for LRGB stars) results in very similar result.

It is noteworthy that the positive lithium-metallicity slope observed in unevolved stars is not preserved in the LRGB phase.
This can be understood as a consequence of lithium depletion during the First Dredge-Up (FDU) process.
Stellar evolution models\citep[e.g.,][]{1990ApJS...73...21D, 1999ApJ...510..217S, 2012MNRAS.419.2195M} indicate that the FDU dilution is metallicity-dependent, which tends to smooth out the initial slope.
Thus, the depletion during stellar evolution may lead to the constant LRGB plateau universally observed, consistent with the findings of Mucciarelli et al. (2022) and this work.

\subsubsection{The Lithium Evolution from LRGB to Highly Evolved RGB} \label{subsubsection:4-3-3}

As a star evolves from the LRGB to a higher stage, lithium becomes more depleted, and its abundance becomes difficult to measure. 
In our sample, 29 stars have been identified as highly evolved RGB stars, and 18 of them do not have detectable lithium. 
In such a situation, we provide only the upper limits for these stars. 
From Figs.~\ref{fig:4} and ~\ref{fig:7}, there is a clear drop in abundance from A(Li) $\sim 1.0$ to A(Li) $< 0.5$ dex although most lithium abundances in highly evolved RGBs are adopted as upper limits. 
A similar rapid depletion has been reported in globular clusters.
\cite{2009A&A...503..545L} demonstrated that lithium depletes abruptly at $M_{\rm v} \sim 3.3$ in NGC\,6397, which corresponds to the luminosity of LRGB in this globular cluster. 
Although the situation in field stars is more complicated, our sample stars exhibit a comparable rapid Li depletion. 
This behavior is consistent with the framework of non-canonical extra mixing. 
It is often triggered near the luminosity-function bump and may be driven by magnetically enhanced thermohaline mixing, rotation-induced meridional circulation, or internal gravity waves\citep{2011ApJ...741...26P,2009PASA...26..168G,2019A&A...621A..24L}. 

In addition, while the abundances for highly evolved stars are close to the detection limits and thus are only available as upper limits, 
these upper limits tend to cluster at similarly low abundance levels, as shown in Fig.~\ref{fig:7}. 
This may reflect the natural evolutionary continuation of lithium depletion from the Spite plateau through the LRGB phase and into the highly evolved RGB stage under sustained extra mixing.
Although the current data are limited by detection thresholds, this pattern may provide a useful reference for future studies of lithium evolution in extremely metal-poor highly evolved stars.

It is worth noting that we cannot rule out the possibility that these highly evolved stars are, or at least part of them, asymptotic giant branch (AGB) stars. Their classification may influence the discussion of Sec.~\ref{subsubsection:4-3-4}.

\subsubsection{Li Abundance in Horizontal Branch Stars}
\label{subsubsection:4-3-4}
The HB stars in our sample can be seen in the top-left panel of Fig.~\ref{fig:6}. These stars are marked with a thick cross in Figs.~\ref{fig:4} and ~\ref{fig:7}.
Due to their higher effective temperatures, HB stars exhibit lithium lines that are similarly weak as those in highly evolved RGB stars, thus we also take the lithium abundances in HB stars as upper limits except for one super Li-rich HB star.
As shown in Fig.~\ref{fig:7}, the upper limits for HB stars are systematically higher than the values of highly evolved RGB stars by approximately $1.0$ dex. However, this comparison involves two sets of upper limits and should therefore be interpreted with caution. To confirm this apparent difference, further studies for HB stars with better signal-to-noise spectra are desirable.

\subsection{Li-rich VMP stars} \label{subsection:4-4}
The anomalous lithium enhancement in low-mass metal-poor stars has been investigated by previous work \cite[e.g.,][]{2018ApJ...852L..31L, 2025A&A...699A.171M}. These Li-rich stars are distributed across a relatively wide range of evolutionary stages, from main-sequence stars to red giant branch stars, but their anomalous surface A(Li) cannot be explained by current stellar evolution models.

In our investigated sample, four common Li-rich stars are identified. These four stars are at different evolutionary stages: one is a turnoff star (J0626+6032), two are RGB stars (J0705+2552 and J0714+1600), and one is an HB star (J0554+5235). These stars exhibit varying degrees of lithium enhancement. Specifically, the NLTE A(Li) of the RGB J0714+1600 is 2.34\,dex, while that of the other three stars exceeds the lithium abundance of the initial solar-system nebula (ISSN). 
J0554=3.72
J0626=3.54
J0705=3.34

\begin{figure}
\centering
\includegraphics[width=0.47\textwidth]{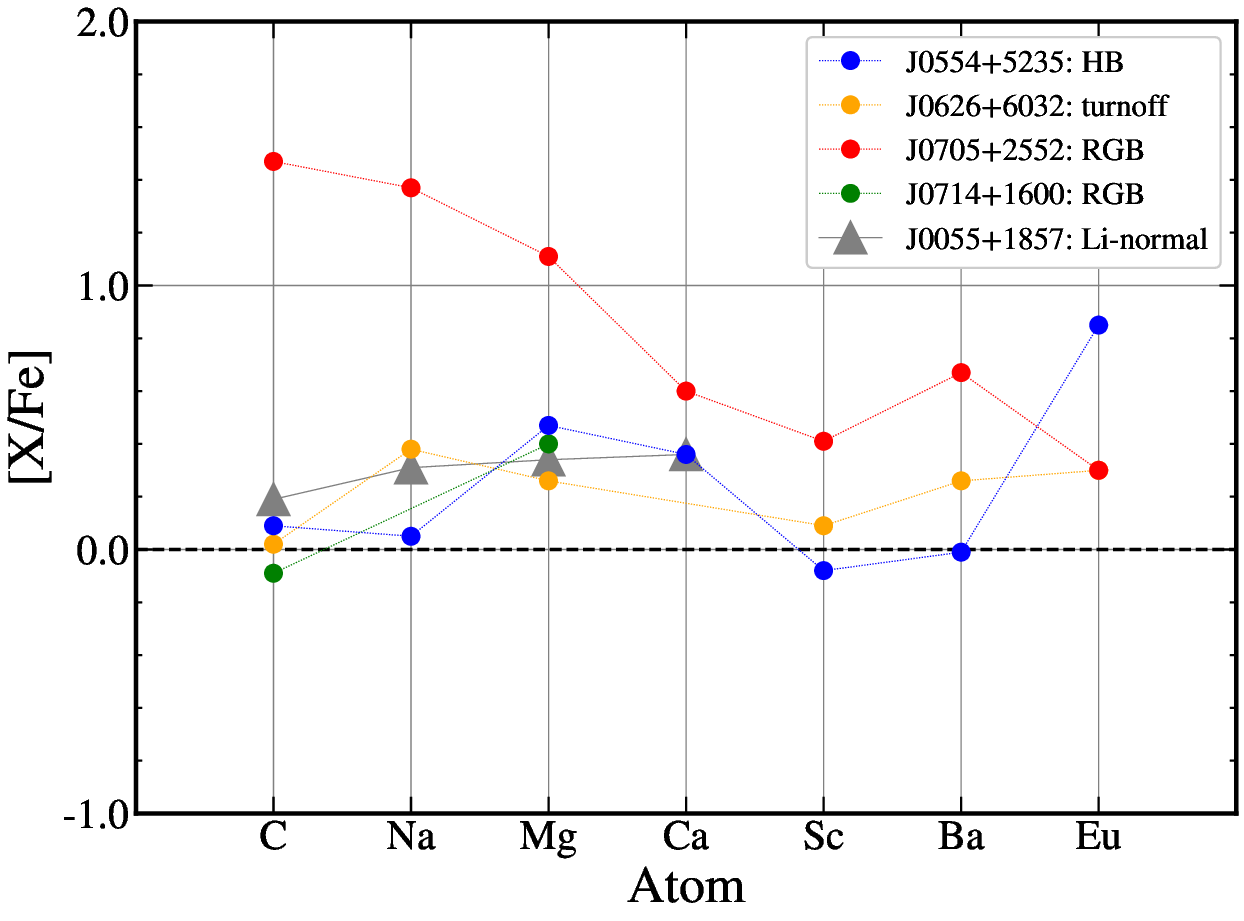}
    \caption{The elemental abundances of four Li-rich stars {in our sample} and one Li-normal star with a similar [Fe/H] in our sample as a comparison. The abundances are presented using NLTE results. 
    For [Eu/Fe] of J0626+6032 and J0705+2552, a limiting value of 0.3~dex is applied. The solar abundances are adopted from \cite{2021A&A...653A.141A}.}
\label{fig:8}
\end{figure}

Fig.~\ref{fig:8} presents their abundance patterns, with results derived from the NLTE analysis in follow-up articles in this series. For the turnoff star J0626+6032 with A(Li)=3.54, its Na abundance shows a slight enhancement but remains within the normal range, while the abundance of other elements of this star does not show any enhancement. For the HB star J0554+5235 with A(Li)=3.72, it shows a normal slight Mg enhancement but also exhibits a Eu enhancement. The abundance patterns of the two stars are completely different. For the LRGB star J0714+1600 with A(Li)=2.34, it only shows a small amount of Mg enhancement, while for the other LRGB star J0705+2552 with A(Li)=3.34, it shows strong C, Mg, and Ba enhancements. 
J0705+2552 has been identified as a carbon enhanced metal-poor (CEMP) star. All of these stars show different abundance patterns compared to the one reported by \cite{2025A&A...699A.171M}.

The mechanism for lithium enhancement is complex and is not yet fully understood. Hypernovae are considered as a potential source to produce lithium, as their high-energy environments can induce spallation reactions of CNO nucli \citep{2002ApJ...581..389F, 2004ApJ...610..888N}. \cite{2025A&A...697A..75M} found the metal-poor star BPM~3066 to be strongly enriched in both Li and Be, with a $^7$Li/$^9$Be ratio of 7.9 matching spallation yields. However, co-enhancements of Li and Na may contradict hypernova predictions. In our Li-rich sample, J0626+6032 shows a moderate Na enhancement. While J0705+2552 has exceptionally high Li and Na enrichment (also see \cite{2018ApJ...852L..31L}). Two other VMP stars have been reported to show similar Li and Na enrichment \citep{2014ApJ...784..158R, 2024A&A...689A..89M}.

Another possible mechanism that has been proposed to explain lithium enhancement is the engulfment of planetary material or a brown dwarf companion \citep{1967Obs....87..238A, 2005ESASP.560..403A}.

Although the engulfment mechanism faces two challenges in fully explaining the high Li abundances observed in these stars. First, the maximum A(Li) after ingestion is usually below 2.2~dex for giants \citep{2016ApJ...833L..24A}. Second, the occurrence rate of planets around metal-poor stars has been debated, with early studies suggesting that this rate is low
\citep{2011ApJ...738L..29K, 2018ApJ...860..109G}, which would reduce the likelihood that all four Li-rich stars would experience engulfment events. However, as demonstrated by \cite{2025A&A...697A..75M}, planetary number may be substantial for stars with [Fe/H]$<-1$\,dex, and observations have confirmed planets around stars with metallicities as low as [Fe/H]$\sim-2$\,dex \citep{2014MNRAS.444..600L}. Three stars (J0554+5235, J0626+6032, J0714+1600) in our Li-rich sample have metallicities similar to that of the confirmed planet-hosting star. While J0705+2552 has [Fe/H] below $\sim -3$\,dex, where planet occurrence remains uncertain. Thus engulfment remains plausible but unconfirmed for this star. 

Among the four Li-rich stars, J0705+2552 shows the most distinctive abundance pattern. This star shows strong C, Na, Mg, and Ba enhancements.
Given its [Ba/Fe] ratio of 0.67\,dex which is close to the commonly adopted threshold, we tentatively classify it as a CEMP-s star. As shown in the lower left panel of Figure~\ref{fig:6}, this star has already completed its FDU phase, and its lithium content is expected to be diluted due to the mixing of lithium-poor interior material. However, more evolved AGB stars can produce Li, Na and $s$-process elements \citep{1971ApJ...164..111C, 2011MNRAS.410.2760V}. The most plausible mechanism for this star is mass transfer from an AGB companion in a binary system. This star needs further investigation. 

In addition, according to \cite{2021NatAs...5...86Y}, the asteroseismic analysis reveals that most Li-rich giants are in the red clump phase, and their lithium abundances are on average much higher than those of Li-rich RGB stars. Among these four Li-rich stars, the HB star J0554+5253 has the highest lithium abundance, which is consistent with the conclusion of \citet{2021NatAs...5...86Y}. This star may have undergone the helium flash, and this event is postulated as the plausible mechanism responsible for its Li enrichment. However, recent studies \citep{2021ApJ...919L...3Z} presented that only He-flash contributes limited Li production, suggesting that additional physical mechanisms are required to trigger abnormal Li enhancement. RGB mergers with a helium white dwarf star (HeWD) may provide possible Li-rich explanation \citep{2020ApJ...889...33Z}. While it remains uncertain whether this mechanism can be applied to the VMP regime. 

The atomic diffusion model is considered to be an effective mechanism for enhancing the lithium content of unevolved stars with temperatures between 6900 and 7100~K \citep{2002ApJ...577L..39D}. However, $\teff$ of the turnoff star J0626+6032 falls far outside this range, so the atomic diffusion mechanism is unlikely to be the cause of its lithium enhancement. In addition to lithium, this star shows only a slight normal enhancement of Na, while it does not show any other elemental enhancement. Lithium can be produced by outward shock waves or neutrino processes in supernovae \citep{1990ApJ...356..272W, 1995ApJS..101..181W}, while the production of lithium in type II supernova explosions is usually accompanied by strong $\alpha$-element enhancements. In binary systems, stars can increase their surface lithium content by accreting material from a companion star, especially when the companion is a red giant, AGB star, or nova. Another possible mechanism is the accretion of circumstellar material rich in lithium, which could be ejected by the nova explosion \citep{2025A&A...699A.171M}.

\section{Summary} \label{section:summary}

In this paper, we present a homogeneous NLTE abundances analysis of lithium for 103 very metal-poor stars with metallicity down to $-4.3$ dex. The sample was selected from the LAMOST survey and observed by the high-resolution spectroscopy of Subaru. The stellar parameters were determined using NLTE line formation calculations for \ion{Fe}{1} and \ion{Fe}{2}. Based on our data, we present new insight into the lithium abundances and lithium plateaus in very/extremely metal-poor stars.

1) The Spite plateau lies at $\sim 2.3$\,dex for warm halo stars. However, it may not be strictly flat, showing a slight positive slope. Most significantly, it appears to extend to lower metallicities as previously suggested, calling into question the reality of the so-called 'meltdown' at low metallicity. The classic Spite plateau stars and the stars in meltdown region have a similar positive slope, indicating a metallicity-extended Spite plateau. 

2) The lithium plateau in LRGB stars was reported by previous work. We confirm that LRGB have a mean lithium abundance of A(Li)$= 1.13$\,dex, which is consistent with previous findings. 

3) The lithium abundance rapidly drops from A(Li) $\sim 1.0$ to A(Li)$<0.5$ as stars evolve from LRGB to highly evolved RGB stars. This may reflect the natural evolutionary continuation of lithium depletion from the Spite plateau through the LRGB phase and into the highly evolved RGB stage under sustained extra mixing. 

4) We report that the upper limit of lithium abundance in horizontal branch stars is possibly higher than that in highly evolved RGB stars.

5) We identify four Li-rich stars in our sample across different evolutionary stages, showing complex and multiple Li production mechanisms. 

Our results suggest that the observed lithium abundances in VMP/EMP stars result from a complex interplay between depletion and production processes.

\begin{acknowledgments}
\noindent\textbf{Acknowledgments} 

This work was supported by the National Natural Science Foundation of China under grant No. 12588202, the National Key R\&D Programs of China No. 2024YFA1611903, the Strategic Priority Research Program of Chinese Academy of Sciences under grant No. XDB1160101, and the and the International Partnership Program of Chinese Academy of Sciences under grant No. 178GJHZ2022040GC. This work was also supported by the National Natural Science Foundation of China under grant Nos. 12222305, 12373036, and 12273055. H.Y. and H.L. acknowledges supports from the Youth Innovation Promotion Association of the CAS. S.A. acknowledges support from the National Natural Science Foundation of China grant No. W2432009. W.A. was supported by JSPS KAKENHI grant Nos. 21H04499 and 25K01046. This research is based on data collected at Subaru Telescope, which is operated by the National Astronomical Observatory of Japan. We are honored and grateful for the opportunity of observing the Universe from Maunakea, which has the cultural, historical and natural significance in Hawaii. The Guoshoujing Telescope (the Large Sky Area Multi-Object Fiber Spectroscopic Telescope LAMOST) is a National Major Scientific Project built by the Chinese Academy of Sciences. LAMOST is operated and managed by the National Astronomical Observatories, Chinese Academy of Sciences. We thank the anonymous referee for the constructive comments, which improved the scientific accuracy and completeness of this work. 

\end{acknowledgments}

\vspace{5mm}
\facilities{Subaru(HDS), LAMOST}

\bibliographystyle{aasjournal}
\bibliography{manuscript}{}

@ARTICLE{2024A&A...691A.142B,
       author = {{Borisov}, Sviatoslav and {Prantzos}, Nikos and {Charbonnel}, Corinne},
        title = "{Evolution of lithium in the disc of the Galaxy and the role of novae}",
      journal = {\aap},
         year = 2024,
        month = nov,
       volume = {691},
          eid = {A142},
        pages = {A142},
          doi = {10.1051/0004-6361/202451321},
archivePrefix = {arXiv},
       eprint = {2410.01880},
 primaryClass = {astro-ph.GA},
       adsurl = {https://ui.adsabs.harvard.edu/abs/2024A&A...691A.142B}
}

@article{Charbonnel2026,
    author = {{Charbonnel}, C. and {Prantzos}, N.},
    title = "{Lithium as a probe of stellar and galactic physics}", 
    journal = {Annual Review of Astronomy and Astrophysics},
    year = 2026,
    eprint = {2602.17470},
    archivePrefix = {arXiv},
    primaryClass = {astro-ph.SR}
}

@ARTICLE{1990ApJS...73...21D,
       author = {{Deliyannis}, Constantine P. and {Demarque}, Pierre and {Kawaler}, Steven D.},
        title = "{Lithium in Halo Stars from Standard Stellar Evolution}",
      journal = {\apjs},
         year = 1990,
        month = may,
       volume = {73},
        pages = {21},
          doi = {10.1086/191439},
       adsurl = {https://ui.adsabs.harvard.edu/abs/1990ApJS...73...21D}
}

@ARTICLE{1999ApJ...510..217S,
       author = {{Sackmann}, I.-Juliana and {Boothroyd}, Arnold I.},
        title = "{Creation of $^{7}$Li and Destruction of $^{3}$He, $^{9}$Be, $^{10}$B, and $^{11}$B in Low-Mass Red Giants, Due to Deep Circulation}",
      journal = {\apj},
         year = 1999,
        month = jan,
       volume = {510},
       number = {1},
        pages = {217-231},
          doi = {10.1086/306545},
       adsurl = {https://ui.adsabs.harvard.edu/abs/1999ApJ...510..217S}
}

@ARTICLE{2018ApJ...855..102C,
       author = {{Cooke}, Ryan J. and {Pettini}, Max and {Steidel}, Charles C.},
        title = "{One Percent Determination of the Primordial Deuterium Abundance}",
      journal = {\apj},
         year = 2018,
        month = mar,
       volume = {855},
       number = {2},
          eid = {102},
        pages = {102},
          doi = {10.3847/1538-4357/aaab53},
archivePrefix = {arXiv},
       eprint = {1710.11129},
 primaryClass = {astro-ph.CO},
       adsurl = {https://ui.adsabs.harvard.edu/abs/2018ApJ...855..102C}
}

@ARTICLE{2012A&A...542A..67P,
       author = {{Prantzos}, N.},
        title = "{Production and evolution of Li, Be, and B isotopes in the Galaxy}",
      journal = {\aap},
         year = 2012,
        month = jun,
       volume = {542},
          eid = {A67},
        pages = {A67},
          doi = {10.1051/0004-6361/201219043},
archivePrefix = {arXiv},
       eprint = {1203.5662},
 primaryClass = {astro-ph.GA},
       adsurl = {https://ui.adsabs.harvard.edu/abs/2012A&A...542A..67P}
}

@ARTICLE{2020ApJ...889...33Z,
       author = {{Zhang}, Xianfei and {Jeffery}, C. Simon and {Li}, Yaguang and {Bi}, Shaolan},
        title = "{Population Synthesis of Helium White Dwarf-Red Giant Star Mergers and the Formation of Lithium-rich Giants and Carbon Stars}",
      journal = {\apj},
         year = 2020,
        month = jan,
       volume = {889},
       number = {1},
          eid = {33},
        pages = {33},
          doi = {10.3847/1538-4357/ab5e89},
archivePrefix = {arXiv},
       eprint = {2001.05600},
 primaryClass = {astro-ph.SR},
       adsurl = {https://ui.adsabs.harvard.edu/abs/2020ApJ...889...33Z}
}

@ARTICLE{2011MNRAS.410.2760V,
       author = {{Ventura}, P. and {D'Antona}, F.},
        title = "{Hot bottom burning in the envelope of super asymptotic giant branch stars}",
      journal = {\mnras},
         year = 2011,
        month = feb,
       volume = {410},
       number = {4},
        pages = {2760-2766},
          doi = {10.1111/j.1365-2966.2010.17651.x},
       adsurl = {https://ui.adsabs.harvard.edu/abs/2011MNRAS.410.2760V}
}

@ARTICLE{1971ApJ...164..111C,
       author = {{Cameron}, A.~G.~W. and {Fowler}, W.~A.},
        title = "{Lithium and the s-PROCESS in Red-Giant Stars}",
      journal = {\apj},
         year = 1971,
        month = feb,
       volume = {164},
        pages = {111},
          doi = {10.1086/150821},
       adsurl = {https://ui.adsabs.harvard.edu/abs/1971ApJ...164..111C}
}

@ARTICLE{2004ApJ...610..888N,
       author = {{Nakamura}, Ko and {Shigeyama}, Toshikazu},
        title = "{Roles of Supernova Ejecta in Nucleosynthesis of the Light Elements Li, Be, and B}",
      journal = {\apj},
         year = 2004,
        month = aug,
       volume = {610},
       number = {2},
        pages = {888-896},
          doi = {10.1086/421840},
archivePrefix = {arXiv},
       eprint = {astro-ph/0404293},
 primaryClass = {astro-ph},
       adsurl = {https://ui.adsabs.harvard.edu/abs/2004ApJ...610..888N}
}

@ARTICLE{2002ApJ...581..389F,
       author = {{Fields}, Brian D. and {Daigne}, Fr{\'e}d{\'e}ric and {Cass{\'e}}, Michel and {Vangioni-Flam}, Elisabeth},
        title = "{Production of Lithium, Beryllium, and Boron by Hypernovae and the Possible Hypernova-Gamma-Ray Burst Connection}",
      journal = {\apj},
         year = 2002,
        month = dec,
       volume = {581},
       number = {1},
        pages = {389-395},
          doi = {10.1086/343853},
archivePrefix = {arXiv},
       eprint = {astro-ph/0107492},
 primaryClass = {astro-ph},
       adsurl = {https://ui.adsabs.harvard.edu/abs/2002ApJ...581..389F}
}

@ARTICLE{2014ApJ...784..158R,
       author = {{Roederer}, Ian U. and {Preston}, George W. and {Thompson}, Ian B. and {Shectman}, Stephen A. and {Sneden}, Christopher},
        title = "{Neutron-capture Nucleosynthesis in the First Stars}",
      journal = {\apj},
         year = 2014,
        month = apr,
       volume = {784},
       number = {2},
          eid = {158},
        pages = {158},
          doi = {10.1088/0004-637X/784/2/158},
archivePrefix = {arXiv},
       eprint = {1402.4144},
 primaryClass = {astro-ph.SR},
       adsurl = {https://ui.adsabs.harvard.edu/abs/2014ApJ...784..158R}
}

@ARTICLE{2024A&A...689A..89M,
       author = {{Mucciarelli}, A. and {Bonifacio}, P. and {Monaco}, L. and {Salaris}, M. and {Matteuzzi}, M.},
        title = "{The true nature of HE 0057-5959, the most metal-poor, Li-rich star}",
      journal = {\aap},
         year = 2024,
        month = sep,
       volume = {689},
          eid = {A89},
        pages = {A89},
          doi = {10.1051/0004-6361/202449290},
archivePrefix = {arXiv},
       eprint = {2406.17026},
 primaryClass = {astro-ph.SR},
       adsurl = {https://ui.adsabs.harvard.edu/abs/2024A&A...689A..89M}
}

@ARTICLE{2014MNRAS.444..600L,
       author = {{Li}, L.-J. and {Qian}, S.-B.},
        title = "{Period analysis of two non-Blazhko RRab stars, FN Lyr and V894 Cyg, based on Kepler photometry: evidence of low-mass companions on wider orbits}",
      journal = {\mnras},
         year = 2014,
        month = oct,
       volume = {444},
       number = {1},
        pages = {600-605},
          doi = {10.1093/mnras/stu1344},
       adsurl = {https://ui.adsabs.harvard.edu/abs/2014MNRAS.444..600L}
}

@ARTICLE{2019MNRAS.489.3539G,
       author = {{Grisoni}, V. and {Matteucci}, F. and {Romano}, D. and {Fu}, X.},
        title = "{Evolution of lithium in the Milky Way halo, discs, and bulge}",
      journal = {\mnras},
         year = 2019,
        month = nov,
       volume = {489},
       number = {3},
        pages = {3539-3546},
          doi = {10.1093/mnras/stz2428},
archivePrefix = {arXiv},
       eprint = {1906.09130},
 primaryClass = {astro-ph.GA},
       adsurl = {https://ui.adsabs.harvard.edu/abs/2019MNRAS.489.3539G}
}

@ARTICLE{1998ApJ...499..735L,
       author = {{Lemoine}, Martin and {Vangioni-Flam}, Elisabeth and {Cass{\'e}}, Michel},
        title = "{Galactic Cosmic Rays and the Evolution of Light Elements}",
      journal = {\apj},
         year = 1998,
        month = may,
       volume = {499},
       number = {2},
        pages = {735-745},
          doi = {10.1086/305650},
       adsurl = {https://ui.adsabs.harvard.edu/abs/1998ApJ...499..735L}
}

@ARTICLE{2021A&A...653A..72R,
       author = {{Romano}, D. and {Magrini}, L. and {Randich}, S. and {Casali}, G. and {Bonifacio}, P. and {Jeffries}, R.~D. and {Matteucci}, F. and {Franciosini}, E. and {Spina}, L. and {Guiglion}, G. and {Chiappini}, C. and {Mucciarelli}, A. and {Ventura}, P. and {Grisoni}, V. and {Bellazzini}, M. and {Bensby}, T. and {Bragaglia}, A. and {de Laverny}, P. and {Korn}, A.~J. and {Martell}, S.~L. and {Tautvai{\v{s}}ien{\.{e}}}, G. and {Carraro}, G. and {Gonneau}, A. and {Jofr{\'e}}, P. and {Pancino}, E. and {Smiljanic}, R. and {Vallenari}, A. and {Fu}, X. and {Guti{\'e}rrez Albarr{\'a}n}, M.~L. and {Jim{\'e}nez-Esteban}, F.~M. and {Montes}, D. and {Damiani}, F. and {Bergemann}, M. and {Worley}, C.},
        title = "{The Gaia-ESO Survey: Galactic evolution of lithium from iDR6}",
      journal = {\aap},
         year = 2021,
        month = sep,
       volume = {653},
          eid = {A72},
        pages = {A72},
          doi = {10.1051/0004-6361/202141340},
archivePrefix = {arXiv},
       eprint = {2106.11614},
 primaryClass = {astro-ph.GA},
       adsurl = {https://ui.adsabs.harvard.edu/abs/2021A&A...653A..72R}
}

@ARTICLE{1975A&A....42...55A,
       author = {{Arnould}, M. and {Norgaard}, H.},
        title = "{The Explosive Thermonuclear Formation of 7Li and 11B}",
      journal = {\aap},
         year = 1975,
        month = aug,
       volume = {42},
        pages = {55},
       adsurl = {https://ui.adsabs.harvard.edu/abs/1975A&A....42...55A}
}

@ARTICLE{1998ApJ...494..680J,
       author = {{Jos{\'e}}, Jordi and {Hernanz}, Margarita},
        title = "{Nucleosynthesis in Classical Novae: CO versus ONe White Dwarfs}",
      journal = {\apj},
         year = 1998,
        month = feb,
       volume = {494},
       number = {2},
        pages = {680-690},
          doi = {10.1086/305244},
archivePrefix = {arXiv},
       eprint = {astro-ph/9709153},
 primaryClass = {astro-ph},
       adsurl = {https://ui.adsabs.harvard.edu/abs/1998ApJ...494..680J}
}

@ARTICLE{1978ApJ...222..600S,
       author = {{Starrfield}, S. and {Truran}, J.~W. and {Sparks}, W.~M. and {Arnould}, M.},
        title = "{On $^{7}$Li production in nova explosions.}",
      journal = {\apj},
         year = 1978,
        month = jun,
       volume = {222},
        pages = {600-603},
          doi = {10.1086/156175},
       adsurl = {https://ui.adsabs.harvard.edu/abs/1978ApJ...222..600S}
}

@ARTICLE{2015Natur.518..381T,
       author = {{Tajitsu}, Akito and {Sadakane}, Kozo and {Naito}, Hiroyuki and {Arai}, Akira and {Aoki}, Wako},
        title = "{Explosive lithium production in the classical nova V339 Del (Nova Delphini 2013)}",
      journal = {\nat},
         year = 2015,
        month = feb,
       volume = {518},
       number = {7539},
        pages = {381-384},
          doi = {10.1038/nature14161},
archivePrefix = {arXiv},
       eprint = {1502.05598},
 primaryClass = {astro-ph.SR},
       adsurl = {https://ui.adsabs.harvard.edu/abs/2015Natur.518..381T}
}

@ARTICLE{1996ApJ...465L..27H,
       author = {{Hernanz}, Margarita and {Jose}, Jordi and {Coc}, Alain and {Isern}, Jordi},
        title = "{On the Synthesis of 7Li and 7Be in Novae}",
      journal = {\apjl},
         year = 1996,
        month = jul,
       volume = {465},
        pages = {L27},
          doi = {10.1086/310122},
       adsurl = {https://ui.adsabs.harvard.edu/abs/1996ApJ...465L..27H}
}

@ARTICLE{2015ApJ...808L..14I,
       author = {{Izzo}, Luca and {Della Valle}, Massimo and {Mason}, Elena and {Matteucci}, Francesca and {Romano}, Donatella and {Pasquini}, Luca and {Vanzi}, Leonardo and {Jordan}, Andres and {Fernandez}, Jos{\'e} Miguel and {Bluhm}, Paz and {Brahm}, Rafael and {Espinoza}, Nestor and {Williams}, Robert},
        title = "{Early Optical Spectra of Nova V1369 Cen Show the Presence of Lithium}",
      journal = {\apjl},
         year = 2015,
        month = jul,
       volume = {808},
       number = {1},
          eid = {L14},
        pages = {L14},
          doi = {10.1088/2041-8205/808/1/L14},
archivePrefix = {arXiv},
       eprint = {1506.08048},
 primaryClass = {astro-ph.SR},
       adsurl = {https://ui.adsabs.harvard.edu/abs/2015ApJ...808L..14I}
}

@ARTICLE{2014MNRAS.442.2058D,
       author = {{Denissenkov}, P.~A. and {Truran}, J.~W. and {Pignatari}, M. and {Trappitsch}, R. and {Ritter}, C. and {Herwig}, F. and {Battino}, U. and {Setoodehnia}, K. and {Paxton}, B.},
        title = "{MESA and NuGrid simulations of classical novae: CO and ONe nova nucleosynthesis}",
      journal = {\mnras},
         year = 2014,
        month = aug,
       volume = {442},
       number = {3},
        pages = {2058-2074},
          doi = {10.1093/mnras/stu1000},
archivePrefix = {arXiv},
       eprint = {1303.6265},
 primaryClass = {astro-ph.SR},
       adsurl = {https://ui.adsabs.harvard.edu/abs/2014MNRAS.442.2058D}
}

@ARTICLE{2016MNRAS.463L.117M,
       author = {{Molaro}, P. and {Izzo}, L. and {Mason}, E. and {Bonifacio}, P. and {Della Valle}, M.},
        title = "{Highly enriched $^{7}$Be in the ejecta of Nova Sagittarii 2015 No. 2 (V5668 Sgr) and the Galactic $^{7}$Li origin}",
      journal = {\mnras},
         year = 2016,
        month = nov,
       volume = {463},
       number = {1},
        pages = {L117-L121},
          doi = {10.1093/mnrasl/slw169},
archivePrefix = {arXiv},
       eprint = {1609.07297},
 primaryClass = {astro-ph.GA},
       adsurl = {https://ui.adsabs.harvard.edu/abs/2016MNRAS.463L.117M}
}

@ARTICLE{2018MNRAS.481.2261S,
       author = {{Selvelli}, P. and {Molaro}, P. and {Izzo}, L.},
        title = "{Absorption and emission features of $^{7}$Be II in the outburst spectra of V838 Her (Nova Her 1991)}",
      journal = {\mnras},
         year = 2018,
        month = dec,
       volume = {481},
       number = {2},
        pages = {2261-2272},
          doi = {10.1093/mnras/sty2310},
archivePrefix = {arXiv},
       eprint = {1809.04180},
 primaryClass = {astro-ph.SR},
       adsurl = {https://ui.adsabs.harvard.edu/abs/2018MNRAS.481.2261S}
}

@ARTICLE{2024ApJ...962..191S,
       author = {{Starrfield}, Sumner and {Bose}, Maitrayee and {Iliadis}, Christian and {Hix}, W. Raphael and {Woodward}, Charles E. and {Wagner}, R. Mark},
        title = "{Hydrodynamic Simulations of Oxygen{\textendash}Neon Classical Novae as Galactic $^{7}$Li Producers and Potential Accretion-induced Collapse Progenitors}",
      journal = {\apj},
         year = 2024,
        month = feb,
       volume = {962},
       number = {2},
          eid = {191},
        pages = {191},
          doi = {10.3847/1538-4357/ad1836},
archivePrefix = {arXiv},
       eprint = {2401.02307},
 primaryClass = {astro-ph.SR},
       adsurl = {https://ui.adsabs.harvard.edu/abs/2024ApJ...962..191S}
}

@ARTICLE{2022ApJ...933L..30K,
       author = {{Kemp}, Alex J. and {Karakas}, Amanda I. and {Casey}, Andrew R. and {C{\^o}t{\'e}}, Benoit and {Izzard}, Robert G. and {Osborn}, Zara},
        title = "{Viability of Novae as Sources of Galactic Lithium}",
      journal = {\apjl},
         year = 2022,
        month = jul,
       volume = {933},
       number = {2},
          eid = {L30},
        pages = {L30},
          doi = {10.3847/2041-8213/ac7c72},
archivePrefix = {arXiv},
       eprint = {2206.13729},
 primaryClass = {astro-ph.SR},
       adsurl = {https://ui.adsabs.harvard.edu/abs/2022ApJ...933L..30K}
}

@ARTICLE{2022MNRAS.509.3258M,
       author = {{Molaro}, P. and {Izzo}, L. and {D'Odorico}, V. and {Aydi}, E. and {Bonifacio}, P. and {Cescutti}, G. and {Harvey}, E.~J. and {Hernanz}, M. and {Selvelli}, P. and {della Valle}, M.},
        title = "{$^{7}$Be in the outburst of the ONe nova V6595 Sgr}",
      journal = {\mnras},
         year = 2022,
        month = jan,
       volume = {509},
       number = {3},
        pages = {3258-3267},
          doi = {10.1093/mnras/stab3106},
archivePrefix = {arXiv},
       eprint = {2111.01469},
 primaryClass = {astro-ph.SR},
       adsurl = {https://ui.adsabs.harvard.edu/abs/2022MNRAS.509.3258M}
}

@ARTICLE{2021ApJ...916...44A,
       author = {{Arai}, Akira and {Tajitsu}, Akito and {Kawakita}, Hideyo and {Shinnaka}, Yoshiharu},
        title = "{Detection of $^{7}$Be II in the Classical Nova V5669 Sgr (Nova Sagittarii 2015 No.3)}",
      journal = {\apj},
         year = 2021,
        month = jul,
       volume = {916},
       number = {1},
          eid = {44},
        pages = {44},
          doi = {10.3847/1538-4357/ac00bf},
archivePrefix = {arXiv},
       eprint = {2106.13448},
 primaryClass = {astro-ph.SR},
       adsurl = {https://ui.adsabs.harvard.edu/abs/2021ApJ...916...44A}
}

@ARTICLE{2024ApJ...971....4G,
       author = {{Gao}, Jun and {Zhu}, Chunhua and {L{\"u}}, Guoliang and {Yu}, Jinlong and {Li}, Lin and {Liu}, Helei and {Guo}, Sufen},
        title = "{Novae: An Important Source of Lithium in the Galaxy}",
      journal = {\apj},
         year = 2024,
        month = aug,
       volume = {971},
       number = {1},
          eid = {4},
        pages = {4},
          doi = {10.3847/1538-4357/ad5a10},
archivePrefix = {arXiv},
       eprint = {2406.13986},
 primaryClass = {astro-ph.SR},
       adsurl = {https://ui.adsabs.harvard.edu/abs/2024ApJ...971....4G}
}

@ARTICLE{2021A&A...653A.141A,
       author = {{Asplund}, M. and {Amarsi}, A.~M. and {Grevesse}, N.},
        title = "{The chemical make-up of the Sun: A 2020 vision}",
      journal = {\aap},
         year = 2021,
        month = sep,
       volume = {653},
          eid = {A141},
        pages = {A141},
          doi = {10.1051/0004-6361/202140445},
archivePrefix = {arXiv},
       eprint = {2105.01661},
 primaryClass = {astro-ph.SR},
       adsurl = {https://ui.adsabs.harvard.edu/abs/2021A&A...653A.141A}
}

@ARTICLE{2019A&A...621A..24L,
       author = {{Lagarde}, N. and {Reyl{\'e}}, C. and {Robin}, A.~C. and {Tautvai{\v{s}}ien{\.{e}}}, G. and {Drazdauskas}, A. and {Mikolaitis}, {\v{S}}. and {Minkevi{\v{c}}i{\={u}}t{\.{e}}}, R. and {Stonkut{\.{e}}}, E. and {Chorniy}, Y. and {Bagdonas}, V. and {Miglio}, A. and {Nasello}, G. and {Gilmore}, G. and {Randich}, S. and {Bensby}, T. and {Bragaglia}, A. and {Flaccomio}, E. and {Francois}, P. and {Korn}, A.~J. and {Pancino}, E. and {Smiljanic}, R. and {Bayo}, A. and {Carraro}, G. and {Costado}, M.~T. and {Jim{\'e}nez-Esteban}, F. and {Jofr{\'e}}, P. and {Martell}, S.~L. and {Masseron}, T. and {Monaco}, L. and {Morbidelli}, L. and {Sbordone}, L. and {Sousa}, S.~G. and {Zaggia}, S.},
        title = "{The Gaia-ESO Survey: impact of extra mixing on C and N abundances of giant stars}",
      journal = {\aap},
         year = 2019,
        month = jan,
       volume = {621},
          eid = {A24},
        pages = {A24},
          doi = {10.1051/0004-6361/201732433},
archivePrefix = {arXiv},
       eprint = {1806.01868},
 primaryClass = {astro-ph.SR},
       adsurl = {https://ui.adsabs.harvard.edu/abs/2019A&A...621A..24L}
}

@ARTICLE{2009PASA...26..168G,
       author = {{Guandalini}, R. and {Palmerini}, S. and {Busso}, M. and {Uttenthaler}, S.},
        title = "{Extra-Mixing in Luminous Cool Red Giants: Hints from Evolved Stars With and Without Li}",
      journal = {\pasa},
         year = 2009,
        month = sep,
       volume = {26},
       number = {3},
        pages = {168-175},
          doi = {10.1071/AS08063},
archivePrefix = {arXiv},
       eprint = {0905.4458},
 primaryClass = {astro-ph.SR},
       adsurl = {https://ui.adsabs.harvard.edu/abs/2009PASA...26..168G}
}

@ARTICLE{2011ApJ...741...26P,
       author = {{Palmerini}, S. and {Cristallo}, S. and {Busso}, M. and {Abia}, C. and {Uttenthaler}, S. and {Gialanella}, L. and {Maiorca}, E.},
        title = "{Deep Mixing in Evolved Stars. II. Interpreting Li Abundances in Red Giant Branch and Asymptotic Giant Branch Stars}",
      journal = {\apj},
         year = 2011,
        month = nov,
       volume = {741},
       number = {1},
          eid = {26},
        pages = {26},
          doi = {10.1088/0004-637X/741/1/26},
archivePrefix = {arXiv},
       eprint = {1107.2844},
 primaryClass = {astro-ph.SR},
       adsurl = {https://ui.adsabs.harvard.edu/abs/2011ApJ...741...26P}
}

@ARTICLE{2025A&A...697A..75M,
       author = {{Monaco}, L. and {Caffau}, E. and {Molaro}, P. and {Bonifacio}, P. and {Cescutti}, G.},
        title = "{A Sequoia stellar candidate with very high $^{7}$Li and $^{9}$Be}",
      journal = {\aap},
         year = 2025,
        month = may,
       volume = {697},
          eid = {A75},
        pages = {A75},
          doi = {10.1051/0004-6361/202453574},
archivePrefix = {arXiv},
       eprint = {2504.21823},
 primaryClass = {astro-ph.GA},
       adsurl = {https://ui.adsabs.harvard.edu/abs/2025A&A...697A..75M}
}

@ARTICLE{2019ApJ...874L..21A,
       author = {{Aguado}, David S. and {Gonz{\'a}lez Hern{\'a}ndez}, Jonay I. and {Allende Prieto}, Carlos and {Rebolo}, Rafael},
        title = "{Back to the Lithium Plateau with the [Fe/H] < -6 Star J0023+0307}",
      journal = {\apjl},
         year = 2019,
        month = apr,
       volume = {874},
       number = {2},
          eid = {L21},
        pages = {L21},
          doi = {10.3847/2041-8213/ab1076},
archivePrefix = {arXiv},
       eprint = {1904.04892},
 primaryClass = {astro-ph.SR},
       adsurl = {https://ui.adsabs.harvard.edu/abs/2019ApJ...874L..21A}
}

@ARTICLE{2021A&A...654A.170M,
       author = {{Matas Pinto}, A.~M. and {Spite}, M. and {Caffau}, E. and {Bonifacio}, P. and {Sbordone}, L. and {Sivarani}, T. and {Steffen}, M. and {Spite}, F. and {Fran{\c{c}}ois}, P. and {Di Matteo}, P.},
        title = "{The metal-poor end of the Spite plateau. II. Chemical and dynamical investigation}",
      journal = {\aap},
         year = 2021,
        month = oct,
       volume = {654},
          eid = {A170},
        pages = {A170},
          doi = {10.1051/0004-6361/202141288},
archivePrefix = {arXiv},
       eprint = {2110.00243},
 primaryClass = {astro-ph.SR},
       adsurl = {https://ui.adsabs.harvard.edu/abs/2021A&A...654A.170M}
}

@ARTICLE{2002Natur.419..904C,
       author = {{Christlieb}, N. and {Bessell}, M.~S. and {Beers}, T.~C. and {Gustafsson}, B. and {Korn}, A. and {Barklem}, P.~S. and {Karlsson}, T. and {Mizuno-Wiedner}, M. and {Rossi}, S.},
        title = "{A stellar relic from the early Milky Way}",
      journal = {\nat},
         year = 2002,
        month = oct,
       volume = {419},
       number = {6910},
        pages = {904-906},
          doi = {10.1038/nature01142},
archivePrefix = {arXiv},
       eprint = {astro-ph/0211274},
 primaryClass = {astro-ph},
       adsurl = {https://ui.adsabs.harvard.edu/abs/2002Natur.419..904C}
}

@ARTICLE{2025A&A...704A.238C,
       author = {{Caffau}, E. and {Steffen}, M. and {Molaro}, P. and {Bonifacio}, P. and {Christlieb}, N. and {Aguado}, D.~S. and {Gonz{\'a}lez Hern{\'a}ndez}, J.~I. and {Zapatero Osorio}, M.~R. and {Monaco}, L. and {Limongi}, M. and {Chieffi}, A. and {Falla}, A. and {Roberti}, L. and {Gallagher}, A.~J. and {Spite}, M. and {Fran{\c{c}}ois}, P. and {Ludwig}, H.-G. and {Sbordone}, L. and {Lallement}, R. and {Allende Prieto}, C. and {Rebolo}, R. and {Cristiani}, S. and {Cupani}, G. and {D'Odorico}, V. and {Martins}, C.~J.~A.~P. and {Milakovi{\'c}}, D. and {Murphy}, M.~T. and {Nunes}, N.~J. and {Santos}, N.~C. and {Schmidt}, T.~M.},
        title = "{Unveiling the nature of HE 0107─5240: a long period binary CEMP-no star with [Fe/H] of ─5.56}",
      journal = {\aap},
         year = 2025,
        month = dec,
       volume = {704},
          eid = {A238},
        pages = {A238},
          doi = {10.1051/0004-6361/202556891},
archivePrefix = {arXiv},
       eprint = {2510.18498},
 primaryClass = {astro-ph.SR},
       adsurl = {https://ui.adsabs.harvard.edu/abs/2025A&A...704A.238C}
}

@ARTICLE{2017ApJ...847..142E,
       author = {{Ezzeddine}, Rana and {Frebel}, Anna and {Plez}, Bertrand},
        title = "{Ultra-metal-poor Stars: Spectroscopic Determination of Stellar Atmospheric Parameters Using Iron Non-LTE Line Abundances}",
      journal = {\apj},
         year = 2017,
        month = oct,
       volume = {847},
       number = {2},
          eid = {142},
        pages = {142},
          doi = {10.3847/1538-4357/aa8875},
archivePrefix = {arXiv},
       eprint = {1612.06309},
 primaryClass = {astro-ph.SR},
       adsurl = {https://ui.adsabs.harvard.edu/abs/2017ApJ...847..142E}
}

@ARTICLE{2017A&A...597A...6N,
       author = {{Nordlander}, T. and {Amarsi}, A.~M. and {Lind}, K. and {Asplund}, M. and {Barklem}, P.~S. and {Casey}, A.~R. and {Collet}, R. and {Leenaarts}, J.},
        title = "{3D NLTE analysis of the most iron-deficient star, SMSS0313-6708}",
      journal = {\aap},
         year = 2017,
        month = jan,
       volume = {597},
          eid = {A6},
        pages = {A6},
          doi = {10.1051/0004-6361/201629202},
archivePrefix = {arXiv},
       eprint = {1609.07416},
 primaryClass = {astro-ph.SR},
       adsurl = {https://ui.adsabs.harvard.edu/abs/2017A&A...597A...6N}
}

@ARTICLE{2016MNRAS.463.1518A,
       author = {{Amarsi}, A.~M. and {Lind}, K. and {Asplund}, M. and {Barklem}, P.~S. and {Collet}, R.},
        title = "{Non-LTE line formation of Fe in late-type stars - III. 3D non-LTE analysis of metal-poor stars}",
      journal = {\mnras},
         year = 2016,
        month = dec,
       volume = {463},
       number = {2},
        pages = {1518-1533},
          doi = {10.1093/mnras/stw2077},
archivePrefix = {arXiv},
       eprint = {1608.06390},
 primaryClass = {astro-ph.SR},
       adsurl = {https://ui.adsabs.harvard.edu/abs/2016MNRAS.463.1518A}
}

@ARTICLE{2025A&ARv..33....2B,
       author = {{Bonifacio}, Piercarlo and {Caffau}, Elisabetta and {Fran{\c{c}}ois}, Patrick and {Spite}, Monique},
        title = "{The most metal-poor stars}",
      journal = {\aapr},
         year = 2025,
        month = jul,
       volume = {33},
       number = {1},
          eid = {2},
        pages = {2},
          doi = {10.1007/s00159-025-00159-2},
archivePrefix = {arXiv},
       eprint = {2504.06335},
 primaryClass = {astro-ph.GA},
       adsurl = {https://ui.adsabs.harvard.edu/abs/2025A&ARv..33....2B}
}

@ARTICLE{2024A&A...690A..38M,
       author = {{Molaro}, P. and {Bonifacio}, P. and {Cupani}, G. and {Howk}, J.~C.},
        title = "{Extragalactic $^{85}$Rb/$^{87}$Rb and $^{6}$Li/$^{7}$Li ratios in the Small Magellanic Cloud}",
      journal = {\aap},
         year = 2024,
        month = oct,
       volume = {690},
          eid = {A38},
        pages = {A38},
          doi = {10.1051/0004-6361/202449529},
archivePrefix = {arXiv},
       eprint = {2407.10818},
 primaryClass = {astro-ph.CO},
       adsurl = {https://ui.adsabs.harvard.edu/abs/2024A&A...690A..38M}
}

@ARTICLE{2024ApJ...966..174Z,
       author = {{Zhang}, Ruizhi and {Matsuno}, Tadafumi and {Li}, Haining and {Aoki}, Wako and {Xue}, Xiang-Xiang and {Suda}, Takuma and {Zhao}, Gang and {Chen}, Yuqin and {Ishigaki}, Miho N. and {Shi}, Jianrong and {Xing}, Qianfan and {Zhao}, Jingkun},
        title = "{Four-hundred Very Metal-poor Stars Studied with LAMOST and Subaru. III. Dynamically Tagged Groups and Chemodynamical Properties}",
      journal = {\apj},
         year = 2024,
        month = may,
       volume = {966},
       number = {2},
          eid = {174},
        pages = {174},
          doi = {10.3847/1538-4357/ad31a6},
archivePrefix = {arXiv},
       eprint = {2405.02978},
 primaryClass = {astro-ph.GA},
       adsurl = {https://ui.adsabs.harvard.edu/abs/2024ApJ...966..174Z}
}

@ARTICLE{2022ApJ...937...52B,
       author = {{Bandyopadhyay}, Avrajit and {Sivarani}, Thirupathi and {Beers}, Timothy C. and {Susmitha}, A. and {Nayak}, Prasanta K. and {Pandey}, Jeewan C.},
        title = "{Li Distribution, Kinematics, and Detailed Abundance Analysis among Very Metal-poor Stars in the Galactic Halo from the HESP-GOMPA Survey}",
      journal = {\apj},
         year = 2022,
        month = oct,
       volume = {937},
       number = {2},
          eid = {52},
        pages = {52},
          doi = {10.3847/1538-4357/ac8b0f},
archivePrefix = {arXiv},
       eprint = {2208.13912},
 primaryClass = {astro-ph.SR},
       adsurl = {https://ui.adsabs.harvard.edu/abs/2022ApJ...937...52B}
}

@ARTICLE{2022ApJ...931..146A,
       author = {{Aoki}, Wako and {Li}, Haining and {Matsuno}, Tadafumi and {Xing}, Qianfan and {Chen}, Yuqin and {Christlieb}, Norbert and {Honda}, Satoshi and {Ishigaki}, Miho N. and {Shi}, Jianrong and {Suda}, Takuma and {Tominaga}, Nozomu and {Yan}, Hong-Liang and {Zhao}, Jingkun and {Zhao}, Gang},
        title = "{Four-hundred Very Metal-poor Stars Studied with LAMOST and Subaru. I. Survey Design, Follow-up Program, and Binary Frequency}",
      journal = {\apj},
         year = 2022,
        month = jun,
       volume = {931},
       number = {2},
          eid = {146},
        pages = {146},
          doi = {10.3847/1538-4357/ac6515},
archivePrefix = {arXiv},
       eprint = {2203.11505},
 primaryClass = {astro-ph.SR},
       adsurl = {https://ui.adsabs.harvard.edu/abs/2022ApJ...931..146A}
}

@ARTICLE{2022ApJ...931..147L,
       author = {{Li}, Haining and {Aoki}, Wako and {Matsuno}, Tadafumi and {Xing}, Qianfan and {Suda}, Takuma and {Tominaga}, Nozomu and {Chen}, Yuqin and {Honda}, Satoshi and {Ishigaki}, Miho N. and {Shi}, Jianrong and {Zhao}, Jingkun and {Zhao}, Gang},
        title = "{Four-hundred Very Metal-poor Stars Studied with LAMOST and Subaru. II. Elemental Abundances}",
      journal = {\apj},
         year = 2022,
        month = jun,
       volume = {931},
       number = {2},
          eid = {147},
        pages = {147},
          doi = {10.3847/1538-4357/ac6514},
archivePrefix = {arXiv},
       eprint = {2203.11529},
 primaryClass = {astro-ph.SR},
       adsurl = {https://ui.adsabs.harvard.edu/abs/2022ApJ...931..147L}
}

@ARTICLE{2022A&A...661A.153M,
       author = {{Mucciarelli}, A. and {Monaco}, L. and {Bonifacio}, P. and {Salaris}, M. and {Deal}, M. and {Spite}, M. and {Richard}, O.~A. and {Lallement}, R.},
        title = "{Discovery of a thin lithium plateau among metal-poor red giant branch stars}",
      journal = {\aap},
         year = 2022,
        month = may,
       volume = {661},
          eid = {A153},
        pages = {A153},
          doi = {10.1051/0004-6361/202142889},
archivePrefix = {arXiv},
       eprint = {2203.10347},
 primaryClass = {astro-ph.SR},
       adsurl = {https://ui.adsabs.harvard.edu/abs/2022A&A...661A.153M}
}

@ARTICLE{2019MNRAS.482.1204H,
       author = {{Hartwig}, Tilman and {Ishigaki}, Miho N. and {Klessen}, Ralf S. and {Yoshida}, Naoki},
        title = "{Fingerprint of the first stars: multi-enriched extremely metal-poor stars in the TOPoS survey}",
      journal = {\mnras},
         year = 2019,
        month = jan,
       volume = {482},
       number = {1},
        pages = {1204-1210},
          doi = {10.1093/mnras/sty2783},
archivePrefix = {arXiv},
       eprint = {1810.04713},
 primaryClass = {astro-ph.GA},
       adsurl = {https://ui.adsabs.harvard.edu/abs/2019MNRAS.482.1204H}
}

@ARTICLE{2018Natur.563...85H,
       author = {{Helmi}, Amina and {Babusiaux}, Carine and {Koppelman}, Helmer H. and {Massari}, Davide and {Veljanoski}, Jovan and {Brown}, Anthony G.~A.},
        title = "{The merger that led to the formation of the Milky Way's inner stellar halo and thick disk}",
      journal = {\nat},
         year = 2018,
        month = oct,
       volume = {563},
       number = {7729},
        pages = {85-88},
          doi = {10.1038/s41586-018-0625-x},
archivePrefix = {arXiv},
       eprint = {1806.06038},
 primaryClass = {astro-ph.GA},
       adsurl = {https://ui.adsabs.harvard.edu/abs/2018Natur.563...85H}
}

@ARTICLE{2018A&A...612A..65B,
       author = {{Bonifacio}, P. and {Caffau}, E. and {Spite}, M. and {Spite}, F. and {Sbordone}, L. and {Monaco}, L. and {Fran{\c{c}}ois}, P. and {Plez}, B. and {Molaro}, P. and {Gallagher}, A.~J. and {Cayrel}, R. and {Christlieb}, N. and {Klessen}, R.~S. and {Koch}, A. and {Ludwig}, H.-G. and {Steffen}, M. and {Zaggia}, S. and {Abate}, C.},
        title = "{TOPoS. IV. Chemical abundances from high-resolution observations of seven extremely metal-poor stars}",
      journal = {\aap},
         year = 2018,
        month = apr,
       volume = {612},
          eid = {A65},
        pages = {A65},
          doi = {10.1051/0004-6361/201732320},
archivePrefix = {arXiv},
       eprint = {1801.03935},
 primaryClass = {astro-ph.SR},
       adsurl = {https://ui.adsabs.harvard.edu/abs/2018A&A...612A..65B}
}

@ARTICLE{2016RvMP...88a5004C,
       author = {{Cyburt}, Richard H. and {Fields}, Brian D. and {Olive}, Keith A. and {Yeh}, Tsung-Han},
        title = "{Big bang nucleosynthesis: Present status}",
      journal = {Reviews of Modern Physics},
         year = 2016,
        month = jan,
       volume = {88},
       number = {1},
          eid = {015004},
        pages = {015004},
          doi = {10.1103/RevModPhys.88.015004},
archivePrefix = {arXiv},
       eprint = {1505.01076},
 primaryClass = {astro-ph.CO},
       adsurl = {https://ui.adsabs.harvard.edu/abs/2016RvMP...88a5004C}
}

@ARTICLE{2015ApJ...808..148S,
       author = {{Sitnova}, T. and {Zhao}, G. and {Mashonkina}, L. and {Chen}, Y. and {Liu}, F. and {Pakhomov}, Yu. and {Tan}, K. and {Bolte}, M. and {Alexeeva}, S. and {Grupp}, F. and {Shi}, J.-R. and {Zhang}, H.-W.},
        title = "{Systematic Non-LTE Study of the -2.6 < [Fe/H] < 0.2 F and G dwarfs in the Solar Neighborhood. I. Stellar Atmosphere Parameters}",
      journal = {\apj},
         year = 2015,
        month = aug,
       volume = {808},
       number = {2},
          eid = {148},
        pages = {148},
          doi = {10.1088/0004-637X/808/2/148},
archivePrefix = {arXiv},
       eprint = {1506.01621},
 primaryClass = {astro-ph.SR},
       adsurl = {https://ui.adsabs.harvard.edu/abs/2015ApJ...808..148S}
}

@ARTICLE{2015ARA&A..53..631F,
       author = {{Frebel}, Anna and {Norris}, John E.},
        title = "{Near-Field Cosmology with Extremely Metal-Poor Stars}",
      journal = {\araa},
         year = 2015,
        month = aug,
       volume = {53},
        pages = {631-688},
          doi = {10.1146/annurev-astro-082214-122423},
archivePrefix = {arXiv},
       eprint = {1501.06921},
 primaryClass = {astro-ph.SR},
       adsurl = {https://ui.adsabs.harvard.edu/abs/2015ARA&A..53..631F}
}

@ARTICLE{2013ARA&A..51..457N,
       author = {{Nomoto}, Ken'ichi and {Kobayashi}, Chiaki and {Tominaga}, Nozomu},
        title = "{Nucleosynthesis in Stars and the Chemical Enrichment of Galaxies}",
      journal = {\araa},
         year = 2013,
        month = aug,
       volume = {51},
       number = {1},
        pages = {457-509},
          doi = {10.1146/annurev-astro-082812-140956},
       adsurl = {https://ui.adsabs.harvard.edu/abs/2013ARA&A..51..457N}
}

@ARTICLE{2010A&A...522A..32R,
       author = {{Romano}, D. and {Karakas}, A.~I. and {Tosi}, M. and {Matteucci}, F.},
        title = "{Quantifying the uncertainties of chemical evolution studies. II. Stellar yields}",
      journal = {\aap},
         year = 2010,
        month = nov,
       volume = {522},
          eid = {A32},
        pages = {A32},
          doi = {10.1051/0004-6361/201014483},
archivePrefix = {arXiv},
       eprint = {1006.5863},
 primaryClass = {astro-ph.GA},
       adsurl = {https://ui.adsabs.harvard.edu/abs/2010A&A...522A..32R}
}

@ARTICLE{2007A&A...462..851B,
       author = {{Bonifacio}, P. and {Molaro}, P. and {Sivarani}, T. and {Cayrel}, R. and {Spite}, M. and {Spite}, F. and {Plez}, B. and {Andersen}, J. and {Barbuy}, B. and {Beers}, T.~C. and {Depagne}, E. and {Hill}, V. and {Fran{\c{c}}ois}, P. and {Nordstr{\"o}m}, B. and {Primas}, F.},
        title = "{First stars VII - Lithium in extremely metal poor dwarfs}",
      journal = {\aap},
         year = 2007,
        month = feb,
       volume = {462},
       number = {3},
        pages = {851-864},
          doi = {10.1051/0004-6361:20064834},
archivePrefix = {arXiv},
       eprint = {astro-ph/0610245},
 primaryClass = {astro-ph},
       adsurl = {https://ui.adsabs.harvard.edu/abs/2007A&A...462..851B}
}

@ARTICLE{2005ARA&A..43..531B,
       author = {{Beers}, Timothy C. and {Christlieb}, Norbert},
        title = "{The Discovery and Analysis of Very Metal-Poor Stars in the Galaxy}",
      journal = {\araa},
         year = 2005,
        month = sep,
       volume = {43},
       number = {1},
        pages = {531-580},
          doi = {10.1146/annurev.astro.42.053102.134057},
       adsurl = {https://ui.adsabs.harvard.edu/abs/2005ARA&A..43..531B}
}

@ARTICLE{2004A&A...416.1117C,
       author = {{Cayrel}, R. and {Depagne}, E. and {Spite}, M. and {Hill}, V. and {Spite}, F. and {Fran{\c{c}}ois}, P. and {Plez}, B. and {Beers}, T. and {Primas}, F. and {Andersen}, J. and {Barbuy}, B. and {Bonifacio}, P. and {Molaro}, P. and {Nordstr{\"o}m}, B.},
        title = "{First stars V - Abundance patterns from C to Zn and supernova yields in the early Galaxy}",
      journal = {\aap},
         year = 2004,
        month = mar,
       volume = {416},
        pages = {1117-1138},
          doi = {10.1051/0004-6361:20034074},
archivePrefix = {arXiv},
       eprint = {astro-ph/0311082},
 primaryClass = {astro-ph},
       adsurl = {https://ui.adsabs.harvard.edu/abs/2004A&A...416.1117C}
}

@ARTICLE{2004ApJ...600..544C,
       author = {{Coc}, Alain and {Vangioni-Flam}, Elisabeth and {Descouvemont}, Pierre and {Adahchour}, Abderrahim and {Angulo}, Carmen},
        title = "{Updated Big Bang Nucleosynthesis Compared with Wilkinson Microwave Anisotropy Probe Observations and the Abundance of Light Elements}",
      journal = {\apj},
         year = 2004,
        month = jan,
       volume = {600},
       number = {2},
        pages = {544-552},
          doi = {10.1086/380121},
archivePrefix = {arXiv},
       eprint = {astro-ph/0309480},
 primaryClass = {astro-ph},
       adsurl = {https://ui.adsabs.harvard.edu/abs/2004ApJ...600..544C}
}

@ARTICLE{1995ApJS..101..181W,
       author = {{Woosley}, S.~E. and {Weaver}, Thomas A.},
        title = "{The Evolution and Explosion of Massive Stars. II. Explosive Hydrodynamics and Nucleosynthesis}",
      journal = {\apjs},
         year = 1995,
        month = nov,
       volume = {101},
        pages = {181},
          doi = {10.1086/192237},
       adsurl = {https://ui.adsabs.harvard.edu/abs/1995ApJS..101..181W}
}

@ARTICLE{1982A&A...115..357S,
       author = {{Spite}, F. and {Spite}, M.},
        title = "{Abundance of lithium in unevolved stars and old disk stars : Interpretation and consequences.}",
      journal = {\aap},
         year = 1982,
        month = nov,
       volume = {115},
        pages = {357-366},
       adsurl = {https://ui.adsabs.harvard.edu/abs/1982A&A...115..357S}
}

@ARTICLE{2025A&A...699A.171M,
       author = {{Matsuno}, Tadafumi and {Kemp}, Alex and {Tanikawa}, Ataru and {Shariat}, Cheyanne E. and {El-Badry}, Kareem J. and {Dodd}, Emma and {Helmi}, Amina and {Koch-Hansen}, Andreas J. and {Yamaguchi}, Natsuko and {Yan}, Hongliang},
        title = "{Unevolved Li-rich stars at low metallicity: A possible formation pathway through novae}",
      journal = {\aap},
         year = 2025,
        month = jul,
       volume = {699},
          eid = {A171},
        pages = {A171},
          doi = {10.1051/0004-6361/202554264},
archivePrefix = {arXiv},
       eprint = {2502.18552},
 primaryClass = {astro-ph.SR},
       adsurl = {https://ui.adsabs.harvard.edu/abs/2025A&A...699A.171M}
}

@ARTICLE{2009A&A...503..545L,
       author = {{Lind}, K. and {Primas}, F. and {Charbonnel}, C. and {Grundahl}, F. and {Asplund}, M.},
        title = "{Signatures of intrinsic Li depletion and Li-Na anti-correlation in the metal-poor globular cluster NGC 6397}",
      journal = {\aap},
         year = 2009,
        month = aug,
       volume = {503},
       number = {2},
        pages = {545-557},
          doi = {10.1051/0004-6361/200912524},
archivePrefix = {arXiv},
       eprint = {0906.2876},
 primaryClass = {astro-ph.SR},
       adsurl = {https://ui.adsabs.harvard.edu/abs/2009A&A...503..545L}
}

@ARTICLE{1999ApJ...523..654R,
       author = {{Ryan}, Sean G. and {Norris}, John E. and {Beers}, Timothy C.},
        title = "{The Spite Lithium Plateau: Ultrathin but Postprimordial}",
      journal = {\apj},
         year = 1999,
        month = oct,
       volume = {523},
       number = {2},
        pages = {654-677},
          doi = {10.1086/307769},
archivePrefix = {arXiv},
       eprint = {astro-ph/9903059},
 primaryClass = {astro-ph},
       adsurl = {https://ui.adsabs.harvard.edu/abs/1999ApJ...523..654R}
}

@ARTICLE{2017AJ....154...52M,
       author = {{Matsuno}, Tadafumi and {Aoki}, Wako and {Beers}, Timothy C. and {Lee}, Young Sun and {Honda}, Satoshi},
        title = "{High-resolution Spectroscopy of Extremely Metal-poor Stars from SDSS/SEGUE. III. Unevolved Stars with [Fe/H] {\ensuremath{\lesssim}} -3.5}",
      journal = {\aj},
         year = 2017,
        month = aug,
       volume = {154},
       number = {2},
          eid = {52},
        pages = {52},
          doi = {10.3847/1538-3881/aa7a08},
archivePrefix = {arXiv},
       eprint = {1706.04712},
 primaryClass = {astro-ph.SR},
       adsurl = {https://ui.adsabs.harvard.edu/abs/2017AJ....154...52M}
}

@ARTICLE{2010A&A...515L...3M,
       author = {{Mel{\'e}ndez}, J. and {Casagrande}, L. and {Ram{\'\i}rez}, I. and {Asplund}, M. and {Schuster}, W.~J.},
        title = "{Observational evidence for a broken Li Spite plateau and mass-dependent Li depletion}",
      journal = {\aap},
         year = 2010,
        month = jun,
       volume = {515},
          eid = {L3},
        pages = {L3},
          doi = {10.1051/0004-6361/200913047},
archivePrefix = {arXiv},
       eprint = {1005.2944},
 primaryClass = {astro-ph.SR},
       adsurl = {https://ui.adsabs.harvard.edu/abs/2010A&A...515L...3M}
}

@ARTICLE{2012MNRAS.419.2195M,
       author = {{Mucciarelli}, A. and {Salaris}, M. and {Bonifacio}, P.},
        title = "{Giants reveal what dwarfs conceal: Li abundance in lower red giant branch stars as diagnostic of the primordial Li}",
      journal = {\mnras},
         year = 2012,
        month = jan,
       volume = {419},
       number = {3},
        pages = {2195-2205},
          doi = {10.1111/j.1365-2966.2011.19870.x},
archivePrefix = {arXiv},
       eprint = {1109.4589},
 primaryClass = {astro-ph.SR},
       adsurl = {https://ui.adsabs.harvard.edu/abs/2012MNRAS.419.2195M}
}

@ARTICLE{1998A&A...333..219Z,
       author = {{Zhao}, G. and {Butler}, K. and {Gehren}, T.},
        title = "{Non-LTE analysis of neutral magnesium in the solar atmosphere}",
      journal = {\aap},
         year = 1998,
        month = may,
       volume = {333},
        pages = {219-230},
       adsurl = {https://ui.adsabs.harvard.edu/abs/1998A&A...333..219Z}
}

@ARTICLE{2023ApJ...957...10A,
      author = {{Alexeeva}, Sofya and {Wang}, Yu and {Zhao}, Gang and {Wang}, Feng and {Wu}, Yong and {Wang}, Jianguo and {Yan}, Hongliang and {Shi}, Jianrong},
       title = "{NLTE Analysis of Y I and Y II in the Atmospheres of FGK Stars}",
     journal = {\apj},
        year = 2023,
       month = nov,
      volume = {957},
      number = {1},
         eid = {10},
       pages = {10},
         doi = {10.3847/1538-4357/acf5e1},
archivePrefix = {arXiv},
      eprint = {2309.01402},
primaryClass = {astro-ph.SR},
      adsurl = {https://ui.adsabs.harvard.edu/abs/2023ApJ...957...10A}
}

@ARTICLE{2018ApJ...866..153A,
      author = {{Alexeeva}, Sofya and {Ryabchikova}, Tatiana and {Mashonkina}, Lyudmila and {Hu}, Shaoming},
       title = "{NLTE Line Formation for Mg I and Mg II in the Atmospheres of B-A-F-G-K Stars}",
     journal = {\apj},
        year = 2018,
       month = oct,
      volume = {866},
      number = {2},
         eid = {153},
       pages = {153},
         doi = {10.3847/1538-4357/aae1a8},
archivePrefix = {arXiv},
      eprint = {1809.06969},
primaryClass = {astro-ph.SR},
      adsurl = {https://ui.adsabs.harvard.edu/abs/2018ApJ...866..153A}
}

@ARTICLE{2014AstL...40..406A,
      author = {{Alexeeva}, S.~A. and {Pakhomov}, Yu. V. and {Mashonkina}, L.~I.},
       title = "{Non-LTE sodium abundance in galactic thick- and thin-disk red giants}",
     journal = {Astronomy Letters},
        year = 2014,
       month = jul,
      volume = {40},
      number = {7},
       pages = {406-424},
         doi = {10.1134/S1063773714070019},
archivePrefix = {arXiv},
      eprint = {1406.4054},
primaryClass = {astro-ph.SR},
      adsurl = {https://ui.adsabs.harvard.edu/abs/2014AstL...40..406A}
}

@ARTICLE{2018PASJ...70...94A,
       author = {{Aoki}, Wako and {Matsuno}, Tadafumi and {Honda}, Satoshi and {Ishigaki}, Miho N. and {Li}, Haining and {Suda}, Takuma and {Kumar}, Yerra Bharat},
        title = "{LAMOST J221750.59+210437.2: A new member of carbon-enhanced extremely metal-poor stars with excesses of Mg and Si}",
      journal = {\pasj},
         year = 2018,
        month = oct,
       volume = {70},
       number = {5},
          eid = {94},
        pages = {94},
          doi = {10.1093/pasj/psy092},
archivePrefix = {arXiv},
       eprint = {1807.11628},
 primaryClass = {astro-ph.SR},
       adsurl = {https://ui.adsabs.harvard.edu/abs/2018PASJ...70...94A}
}

@ARTICLE{2018ApJ...852L..31L,
       author = {{Li}, Haining and {Aoki}, Wako and {Matsuno}, Tadafumi and {Bharat Kumar}, Yerra and {Shi}, Jianrong and {Suda}, Takuma and {Zhao}, Gang},
        title = "{Enormous Li Enhancement Preceding Red Giant Phases in Low-mass Stars in the Milky Way Halo}",
      journal = {\apjl},
         year = 2018,
        month = jan,
       volume = {852},
       number = {2},
          eid = {L31},
        pages = {L31},
          doi = {10.3847/2041-8213/aaa438},
archivePrefix = {arXiv},
       eprint = {1801.00090},
 primaryClass = {astro-ph.SR},
       adsurl = {https://ui.adsabs.harvard.edu/abs/2018ApJ...852L..31L}
}

@ARTICLE{2007A&A...465..587S,
       author = {{Shi}, J.~R. and {Gehren}, T. and {Zhang}, H.~W. and {Zeng}, J.~L. and {Zhao}, G.},
        title = "{Lithium abundances in metal-poor stars}",
      journal = {\aap},
         year = 2007,
        month = apr,
       volume = {465},
       number = {2},
        pages = {587-591},
          doi = {10.1051/0004-6361:20066709},
       adsurl = {https://ui.adsabs.harvard.edu/abs/2007A&A...465..587S}
}

@ARTICLE{2018NatAs...2..790Y,
       author = {{Yan}, Hong-Liang and {Shi}, Jian-Rong and {Zhou}, Yu-Tao and {Chen}, Yong-Shou and {Li}, Er-Tao and {Zhang}, Suyalatu and {Bi}, Shao-Lan and {Wu}, Ya-Qian and {Li}, Zhi-Hong and {Guo}, Bing and {Liu}, Wei-Ping and {Gao}, Qi and {Zhang}, Jun-Bo and {Zhou}, Ze-Ming and {Li}, Hai-Ning and {Zhao}, Gang},
        title = "{The nature of the lithium enrichment in the most Li-rich giant star}",
      journal = {Nature Astronomy},
         year = 2018,
        month = aug,
       volume = {2},
        pages = {790-795},
          doi = {10.1038/s41550-018-0544-7},
archivePrefix = {arXiv},
       eprint = {1809.00187},
 primaryClass = {astro-ph.SR},
       adsurl = {https://ui.adsabs.harvard.edu/abs/2018NatAs...2..790Y}
}

@ARTICLE{1998MNRAS.296.1057B,
       author = {{Barklem}, P.~S. and {O'Mara}, B.~J. and {Ross}, J.~E.},
        title = "{The broadening of d-f and f-d transitions by collisions with neutral hydrogen atoms}",
      journal = {\mnras},
         year = 1998,
        month = jun,
       volume = {296},
       number = {4},
        pages = {1057-1060},
          doi = {10.1046/j.1365-8711.1998.01484.x},
       adsurl = {https://ui.adsabs.harvard.edu/abs/1998MNRAS.296.1057B}
}

@PHDTHESIS{1999PhDT.......216R,
       author = {{Reetz}, Johannes Krischna},
        title = "{Sauerstoff in k{\"u}hlen Sternen und die chemische Entwicklung der GalaxisSauerstoff in k{\"u}hlen Sternen und die chemische Entwicklung der Galaxis}",
       school = {Ludwig-Maximilians University of Munich, Germany},
         year = 1999,
        month = jan,
       adsurl = {https://ui.adsabs.harvard.edu/abs/1999PhDT.......216R}
}

@ARTICLE{2008A&A...486..951G,
       author = {{Gustafsson}, B. and {Edvardsson}, B. and {Eriksson}, K. and {J{\o}rgensen}, U.~G. and {Nordlund}, {\r{A}}. and {Plez}, B.},
        title = "{A grid of MARCS model atmospheres for late-type stars. I. Methods and general properties}",
      journal = {\aap},
         year = 2008,
        month = aug,
       volume = {486},
       number = {3},
        pages = {951-970},
          doi = {10.1051/0004-6361:200809724},
archivePrefix = {arXiv},
       eprint = {0805.0554},
 primaryClass = {astro-ph},
       adsurl = {https://ui.adsabs.harvard.edu/abs/2008A&A...486..951G}
}

@ARTICLE{2025ApJ...983..127S,
       author = {{Shi}, Jianrong and {Zhao}, Gang and {Liu}, Shuai and {Zhou}, Zeming and {Li}, Haining and {Yan}, Hongliang and {Alexeeva}, Sofya and {Zhang}, Huawei and {Aoki}, Wako and {Matsuno}, Tadafumi and {Zhao}, Jingkun and {Chen}, Huiling and {Shen}, Yufu},
        title = "{A Systematic NLTE Study of Very Metal-poor Stars with Metallicity down to ‑4.3 dex. I. Global Stellar Parameters Based on High-resolution Spectra}",
      journal = {\apj},
         year = 2025,
        month = apr,
       volume = {983},
       number = {2},
          eid = {127},
        pages = {127},
          doi = {10.3847/1538-4357/adc127},
       adsurl = {https://ui.adsabs.harvard.edu/abs/2025ApJ...983..127S}
}

@ARTICLE{1991A&A...245..171R,
       author = {{Rybicki}, G.~B. and {Hummer}, D.~G.},
        title = "{An accelerated lambda iteration method for multilevel radiative transfer. I. Non-overlapping lines with background continuum}",
      journal = {\aap},
         year = 1991,
        month = may,
       volume = {245},
        pages = {171-181},
       adsurl = {https://ui.adsabs.harvard.edu/abs/1991A&A...245..171R}
}

@ARTICLE{1992A&A...262..209R,
       author = {{Rybicki}, G.~B. and {Hummer}, D.~G.},
        title = "{An accelerated lambda iteration method for multilevel radiative transfer. II. Overlapping transitions with full continuum.}",
      journal = {\aap},
         year = 1992,
        month = aug,
       volume = {262},
        pages = {209-215},
       adsurl = {https://ui.adsabs.harvard.edu/abs/1992A&A...262..209R}
}

@ARTICLE{2011A&A...528A..87M,
       author = {{Mashonkina}, L. and {Gehren}, T. and {Shi}, J. -R. and {Korn}, A.~J. and {Grupp}, F.},
        title = "{A non-LTE study of neutral and singly-ionized iron line spectra in 1D models of the Sun and selected late-type stars}",
      journal = {\aap},
         year = 2011,
        month = apr,
       volume = {528},
          eid = {A87},
        pages = {A87},
          doi = {10.1051/0004-6361/201015336},
archivePrefix = {arXiv},
       eprint = {1101.4570},
 primaryClass = {astro-ph.SR},
       adsurl = {https://ui.adsabs.harvard.edu/abs/2011A&A...528A..87M}
}

@ARTICLE{2016ApJ...833..225Z,
       author = {{Zhao}, G. and {Mashonkina}, L. and {Yan}, H.~L. and {Alexeeva}, S. and {Kobayashi}, C. and {Pakhomov}, Yu. and {Shi}, J.~R. and {Sitnova}, T. and {Tan}, K.~F. and {Zhang}, H.~W. and {Zhang}, J.~B. and {Zhou}, Z.~M. and {Bolte}, M. and {Chen}, Y.~Q. and {Li}, X. and {Liu}, F. and {Zhai}, M.},
        title = "{Systematic Non-LTE Study of the -.6 {\ensuremath{\leq}} [Fe/H] {\ensuremath{\leq}} 0.2 F and G Dwarfs in the Solar Neighborhood. II. Abundance Patterns from Li to Eu}",
      journal = {\apj},
         year = 2016,
        month = dec,
       volume = {833},
       number = {2},
          eid = {225},
        pages = {225},
          doi = {10.3847/1538-4357/833/2/225},
archivePrefix = {arXiv},
       eprint = {1610.00193},
 primaryClass = {astro-ph.SR},
       adsurl = {https://ui.adsabs.harvard.edu/abs/2016ApJ...833..225Z}
}

@ARTICLE{2004Ap&SS.291..261Y,
       author = {{Yi}, Sukyoung K. and {demarque}, Pierre and {Kim}, Yong-Cheol},
        title = "{The Y $^{2}$ Isochrones}",
      journal = {\apss},
         year = 2004,
        month = jun,
       volume = {291},
       number = {3},
        pages = {261-262},
          doi = {10.1023/B:ASTR.0000044330.92199.e2},
archivePrefix = {arXiv},
       eprint = {astro-ph/0409024},
 primaryClass = {astro-ph},
       adsurl = {https://ui.adsabs.harvard.edu/abs/2004Ap&SS.291..261Y}
}

@ARTICLE{1967Obs....87..238A,
       author = {{Alexander}, J.~B.},
        title = "{A possible source of lithium in the atmospheres of some red giants}",
      journal = {The Observatory},
         year = 1967,
        month = oct,
       volume = {87},
        pages = {238-240},
       adsurl = {https://ui.adsabs.harvard.edu/abs/1967Obs....87..238A}
}

@INPROCEEDINGS{2005ESASP.560..403A,
       author = {{Ashwell}, J.~F. and {Jeffries}, R.~D. and {Smalley}, B. and {Deliyannis}, C.~P. and {Steinhauer}, A. and {King}, J.~R.},
        title = "{The Li overabundance of J37: Diffusion or accretion?}",
    booktitle = {13th Cambridge Workshop on Cool Stars, Stellar Systems and the Sun},
         year = 2005,
       editor = {{Favata}, F. and {Hussain}, G.~A.~J. and {Battrick}, B.},
       series = {ESA Special Publication},
       volume = {560},
        month = mar,
        pages = {403},
       adsurl = {https://ui.adsabs.harvard.edu/abs/2005ESASP.560..403A}
}

@ARTICLE{2011ApJ...738L..29K,
       author = {{Koch}, Andreas and {Lind}, Karin and {Rich}, R. Michael},
        title = "{Discovery of a Super-Li-rich Turnoff Star in the Metal-poor Globular Cluster NGC 6397}",
      journal = {\apjl},
         year = 2011,
        month = sep,
       volume = {738},
       number = {2},
          eid = {L29},
        pages = {L29},
          doi = {10.1088/2041-8205/738/2/L29},
archivePrefix = {arXiv},
       eprint = {1108.2033},
 primaryClass = {astro-ph.GA},
       adsurl = {https://ui.adsabs.harvard.edu/abs/2011ApJ...738L..29K}
}

@ARTICLE{2018ApJ...860..109G,
       author = {{Ghezzi}, Luan and {Montet}, Benjamin T. and {Johnson}, John Asher},
        title = "{Retired A Stars Revisited: An Updated Giant Planet Occurrence Rate as a Function of Stellar Metallicity and Mass}",
      journal = {\apj},
         year = 2018,
        month = jun,
       volume = {860},
       number = {2},
          eid = {109},
        pages = {109},
          doi = {10.3847/1538-4357/aac37c},
archivePrefix = {arXiv},
       eprint = {1804.09082},
 primaryClass = {astro-ph.SR},
       adsurl = {https://ui.adsabs.harvard.edu/abs/2018ApJ...860..109G}
}

@ARTICLE{2016ApJ...833L..24A,
       author = {{Aguilera-G{\'o}mez}, Claudia and {Chanam{\'e}}, Julio and {Pinsonneault}, Marc H. and {Carlberg}, Joleen K.},
        title = "{On Lithium-rich Red Giants: Engulfment on the Giant Branch of Trumpler 20}",
      journal = {\apjl},
         year = 2016,
        month = dec,
       volume = {833},
       number = {2},
          eid = {L24},
        pages = {L24},
          doi = {10.3847/2041-8213/833/2/L24},
archivePrefix = {arXiv},
       eprint = {1609.07492},
 primaryClass = {astro-ph.SR},
       adsurl = {https://ui.adsabs.harvard.edu/abs/2016ApJ...833L..24A}
}

@ARTICLE{2002ApJ...577L..39D,
       author = {{Deliyannis}, Constantine P. and {Steinhauer}, Aaron and {Jeffries}, R.~D.},
        title = "{Discovery of the Most Lithium-rich Dwarf: Diffusion in Action}",
      journal = {\apjl},
         year = 2002,
        month = sep,
       volume = {577},
       number = {1},
        pages = {L39-L43},
          doi = {10.1086/344046},
       adsurl = {https://ui.adsabs.harvard.edu/abs/2002ApJ...577L..39D}
}

@ARTICLE{1990ApJ...356..272W,
       author = {{Woosley}, S.~E. and {Hartmann}, D.~H. and {Hoffman}, R.~D. and {Haxton}, W.~C.},
        title = "{The nu -Process}",
      journal = {\apj},
         year = 1990,
        month = jun,
       volume = {356},
        pages = {272},
          doi = {10.1086/168839},
       adsurl = {https://ui.adsabs.harvard.edu/abs/1990ApJ...356..272W}
}

@ARTICLE{2021NatAs...5...86Y,
       author = {{Yan}, Hong-Liang and {Zhou}, Yu-Tao and {Zhang}, Xianfei and {Li}, Yaguang and {Gao}, Qi and {Shi}, Jian-Rong and {Zhao}, Gang and {Aoki}, Wako and {Matsuno}, Tadafumi and {Li}, Yan and {Xu}, Xiao-Dong and {Li}, Haining and {Wu}, Ya-Qian and {Jin}, Meng-Qi and {Mosser}, Benoit and {Bi}, Shao-Lan and {Fu}, Jian-Ning and {Pan}, Kaike and {Suda}, Takuma and {Liu}, Yu-Juan and {Zhao}, Jing-Kun and {Liang}, Xi-Long},
        title = "{Most lithium-rich low-mass evolved stars revealed as red clump stars by asteroseismology and spectroscopy}",
      journal = {Nature Astronomy},
         year = 2021,
        month = jan,
       volume = {5},
        pages = {86-93},
          doi = {10.1038/s41550-020-01217-8},
archivePrefix = {arXiv},
       eprint = {2010.02106},
 primaryClass = {astro-ph.SR},
       adsurl = {https://ui.adsabs.harvard.edu/abs/2021NatAs...5...86Y}
}

@ARTICLE{2018MNRAS.475.1537M,
       author = {{Myeong}, G.~C. and {Evans}, N.~W. and {Belokurov}, V. and {Amorisco}, N.~C. and {Koposov}, S.~E.},
        title = "{Halo substructure in the SDSS-Gaia catalogue: streams and clumps}",
      journal = {\mnras},
         year = 2018,
        month = apr,
       volume = {475},
       number = {2},
        pages = {1537-1548},
          doi = {10.1093/mnras/stx3262},
archivePrefix = {arXiv},
       eprint = {1712.04071},
 primaryClass = {astro-ph.GA},
       adsurl = {https://ui.adsabs.harvard.edu/abs/2018MNRAS.475.1537M}
}

@ARTICLE{2020ApJ...895..138K,
       author = {{Kobayashi}, Chiaki and {Leung}, Shing-Chi and {Nomoto}, Ken'ichi},
        title = "{New Type Ia Supernova Yields and the Manganese and Nickel Problems in the Milky Way and Dwarf Spheroidal Galaxies}",
      journal = {\apj},
         year = 2020,
        month = jun,
       volume = {895},
       number = {2},
          eid = {138},
        pages = {138},
          doi = {10.3847/1538-4357/ab8e44},
archivePrefix = {arXiv},
       eprint = {1906.09980},
 primaryClass = {astro-ph.HE},
       adsurl = {https://ui.adsabs.harvard.edu/abs/2020ApJ...895..138K}
}

@ARTICLE{2016A&A...596A.109P,
       author = {{Planck Collaboration} and {Aghanim}, N. and {Ashdown}, M. and {Aumont}, J. and {Baccigalupi}, C. and {Ballardini}, M. and {Banday}, A.~J. and {Barreiro}, R.~B. and {Bartolo}, N. and {Basak}, S. and {Benabed}, K. and {Bernard}, J. -P. and {Bersanelli}, M. and {Bielewicz}, P. and {Bonavera}, L. and {Bond}, J.~R. and {Borrill}, J. and {Bouchet}, F.~R. and {Boulanger}, F. and {Burigana}, C. and {Calabrese}, E. and {Cardoso}, J. -F. and {Carron}, J. and {Chiang}, H.~C. and {Colombo}, L.~P.~L. and {Comis}, B. and {Couchot}, F. and {Coulais}, A. and {Crill}, B.~P. and {Curto}, A. and {Cuttaia}, F. and {de Bernardis}, P. and {de Zotti}, G. and {Delabrouille}, J. and {Di Valentino}, E. and {Dickinson}, C. and {Diego}, J.~M. and {Dor{\'e}}, O. and {Douspis}, M. and {Ducout}, A. and {Dupac}, X. and {Dusini}, S. and {Elsner}, F. and {En{\ss}lin}, T.~A. and {Eriksen}, H.~K. and {Falgarone}, E. and {Fantaye}, Y. and {Finelli}, F. and {Forastieri}, F. and {Frailis}, M. and {Fraisse}, A.~A. and {Franceschi}, E. and {Frolov}, A. and {Galeotta}, S. and {Galli}, S. and {Ganga}, K. and {G{\'e}nova-Santos}, R.~T. and {Gerbino}, M. and {Ghosh}, T. and {Giraud-H{\'e}raud}, Y. and {Gonz{\'a}lez-Nuevo}, J. and {G{\'o}rski}, K.~M. and {Gruppuso}, A. and {Gudmundsson}, J.~E. and {Hansen}, F.~K. and {Helou}, G. and {Henrot-Versill{\'e}}, S. and {Herranz}, D. and {Hivon}, E. and {Huang}, Z. and {Jaffe}, A.~H. and {Jones}, W.~C. and {Keih{\"a}nen}, E. and {Keskitalo}, R. and {Kiiveri}, K. and {Kisner}, T.~S. and {Krachmalnicoff}, N. and {Kunz}, M. and {Kurki-Suonio}, H. and {Lamarre}, J. -M. and {Langer}, M. and {Lasenby}, A. and {Lattanzi}, M. and {Lawrence}, C.~R. and {Le Jeune}, M. and {Levrier}, F. and {Lilje}, P.~B. and {Lilley}, M. and {Lindholm}, V. and {L{\'o}pez-Caniego}, M. and {Ma}, Y. -Z. and {Mac{\'\i}as-P{\'e}rez}, J.~F. and {Maggio}, G. and {Maino}, D. and {Mandolesi}, N. and {Mangilli}, A. and {Maris}, M. and {Martin}, P.~G. and {Mart{\'\i}nez-Gonz{\'a}lez}, E. and {Matarrese}, S. and {Mauri}, N. and {McEwen}, J.~D. and {Melchiorri}, A. and {Mennella}, A. and {Migliaccio}, M. and {Miville-Desch{\^e}nes}, M. -A. and {Molinari}, D. and {Moneti}, A. and {Montier}, L. and {Morgante}, G. and {Moss}, A. and {Natoli}, P. and {Oxborrow}, C.~A. and {Pagano}, L. and {Paoletti}, D. and {Patanchon}, G. and {Perdereau}, O. and {Perotto}, L. and {Pettorino}, V. and {Piacentini}, F. and {Plaszczynski}, S. and {Polastri}, L. and {Polenta}, G. and {Puget}, J. -L. and {Rachen}, J.~P. and {Racine}, B. and {Reinecke}, M. and {Remazeilles}, M. and {Renzi}, A. and {Rocha}, G. and {Rosset}, C. and {Rossetti}, M. and {Roudier}, G. and {Rubi{\~n}o-Mart{\'\i}n}, J.~A. and {Ruiz-Granados}, B. and {Salvati}, L. and {Sandri}, M. and {Savelainen}, M. and {Scott}, D. and {Sirignano}, C. and {Sirri}, G. and {Soler}, J.~D. and {Spencer}, L.~D. and {Suur-Uski}, A. -S. and {Tauber}, J.~A. and {Tavagnacco}, D. and {Tenti}, M. and {Toffolatti}, L. and {Tomasi}, M. and {Tristram}, M. and {Trombetti}, T. and {Valiviita}, J. and {Van Tent}, F. and {Vielva}, P. and {Villa}, F. and {Vittorio}, N. and {Wandelt}, B.~D. and {Wehus}, I.~K. and {Zacchei}, A. and {Zonca}, A.},
        title = "{Planck intermediate results. XLVIII. Disentangling Galactic dust emission and cosmic infrared background anisotropies}",
      journal = {\aap},
         year = 2016,
        month = dec,
       volume = {596},
          eid = {A109},
        pages = {A109},
          doi = {10.1051/0004-6361/201629022},
archivePrefix = {arXiv},
       eprint = {1605.09387},
 primaryClass = {astro-ph.CO},
       adsurl = {https://ui.adsabs.harvard.edu/abs/2016A&A...596A.109P}
}

@ARTICLE{2010A&A...522A..26S,
       author = {{Sbordone}, L. and {Bonifacio}, P. and {Caffau}, E. and {Ludwig}, H. -G. and {Behara}, N.~T. and {Gonz{\'a}lez Hern{\'a}ndez}, J.~I. and {Steffen}, M. and {Cayrel}, R. and {Freytag}, B. and {van't Veer}, C. and {Molaro}, P. and {Plez}, B. and {Sivarani}, T. and {Spite}, M. and {Spite}, F. and {Beers}, T.~C. and {Christlieb}, N. and {Fran{\c{c}}ois}, P. and {Hill}, V.},
        title = "{The metal-poor end of the Spite plateau. I. Stellar parameters, metallicities, and lithium abundances}",
      journal = {\aap},
         year = 2010,
        month = nov,
       volume = {522},
          eid = {A26},
        pages = {A26},
          doi = {10.1051/0004-6361/200913282},
archivePrefix = {arXiv},
       eprint = {1003.4510},
 primaryClass = {astro-ph.GA},
       adsurl = {https://ui.adsabs.harvard.edu/abs/2010A&A...522A..26S}
}

@ARTICLE{2015MNRAS.452.3256F,
       author = {{Fu}, Xiaoting and {Bressan}, Alessandro and {Molaro}, Paolo and {Marigo}, Paola},
        title = "{Lithium evolution in metal-poor stars: from pre-main sequence to the Spite plateau}",
      journal = {\mnras},
         year = 2015,
        month = sep,
       volume = {452},
       number = {3},
        pages = {3256-3265},
          doi = {10.1093/mnras/stv1384},
archivePrefix = {arXiv},
       eprint = {1506.05993},
 primaryClass = {astro-ph.SR},
       adsurl = {https://ui.adsabs.harvard.edu/abs/2015MNRAS.452.3256F}
}

@ARTICLE{2009PhR...472....1I,
       author = {{Iocco}, Fabio and {Mangano}, Gianpiero and {Miele}, Gennaro and {Pisanti}, Ofelia and {Serpico}, Pasquale D.},
        title = "{Primordial nucleosynthesis: From precision cosmology to fundamental physics}",
      journal = {\physrep},
         year = 2009,
        month = mar,
       volume = {472},
       number = {1-6},
        pages = {1-76},
          doi = {10.1016/j.physrep.2009.02.002},
archivePrefix = {arXiv},
       eprint = {0809.0631},
 primaryClass = {astro-ph},
       adsurl = {https://ui.adsabs.harvard.edu/abs/2009PhR...472....1I}
}

@ARTICLE{1997MNRAS.285..847B,
       author = {{Bonifacio}, P. and {Molaro}, P.},
        title = "{The primordial lithium abundance}",
      journal = {\mnras},
         year = 1997,
        month = mar,
       volume = {285},
       number = {4},
        pages = {847-861},
          doi = {10.1093/mnras/285.4.847},
archivePrefix = {arXiv},
       eprint = {astro-ph/9611043},
 primaryClass = {astro-ph},
       adsurl = {https://ui.adsabs.harvard.edu/abs/1997MNRAS.285..847B}
}

@ARTICLE{2009ApJ...698.1803A,
       author = {{Aoki}, Wako and {Barklem}, Paul S. and {Beers}, Timothy C. and {Christlieb}, Norbert and {Inoue}, Susumu and {Garc{\'\i}a P{\'e}rez}, Ana E. and {Norris}, John E. and {Carollo}, Daniela},
        title = "{Lithium Abundances of Extremely Metal-Poor Turnoff Stars}",
      journal = {\apj},
         year = 2009,
        month = jun,
       volume = {698},
       number = {2},
        pages = {1803-1812},
          doi = {10.1088/0004-637X/698/2/1803},
archivePrefix = {arXiv},
       eprint = {0904.1448},
 primaryClass = {astro-ph.SR},
       adsurl = {https://ui.adsabs.harvard.edu/abs/2009ApJ...698.1803A}
}

@ARTICLE{2004A&A...413.1045G,
       author = {{Gehren}, T. and {Liang}, Y.~C. and {Shi}, J.~R. and {Zhang}, H.~W. and {Zhao}, G.},
        title = "{Abundances of Na, Mg and Al in nearby metal-poor stars}",
      journal = {\aap},
         year = 2004,
        month = jan,
       volume = {413},
        pages = {1045-1063},
          doi = {10.1051/0004-6361:20031582},
       adsurl = {https://ui.adsabs.harvard.edu/abs/2004A&A...413.1045G}
}

@ARTICLE{2008A&A...486..303S,
       author = {{Shi}, J.~R. and {Gehren}, T. and {Butler}, K. and {Mashonkina}, L.~I. and {Zhao}, G.},
        title = "{Statistical equilibrium of silicon in the solar atmosphere}",
      journal = {\aap},
         year = 2008,
        month = jul,
       volume = {486},
       number = {1},
        pages = {303-310},
          doi = {10.1051/0004-6361:200809452},
archivePrefix = {arXiv},
       eprint = {0805.3564},
 primaryClass = {astro-ph},
       adsurl = {https://ui.adsabs.harvard.edu/abs/2008A&A...486..303S}
}

@ARTICLE{2015ApJ...802...36Y,
       author = {{Yan}, H.~L. and {Shi}, J.~R. and {Zhao}, G.},
        title = "{Non-LTE Analysis of Neutral Copper in Late-type Metal-poor Stars}",
      journal = {\apj},
         year = 2015,
        month = mar,
       volume = {802},
       number = {1},
          eid = {36},
        pages = {36},
          doi = {10.1088/0004-637X/802/1/36},
archivePrefix = {arXiv},
       eprint = {1503.06269},
 primaryClass = {astro-ph.SR},
       adsurl = {https://ui.adsabs.harvard.edu/abs/2015ApJ...802...36Y}
}

@ARTICLE{2021MNRAS.500.2159W,
       author = {{Wang}, Ella Xi and {Nordlander}, Thomas and {Asplund}, Martin and {Amarsi}, Anish M. and {Lind}, Karin and {Zhou}, Yixiao},
        title = "{3D NLTE spectral line formation of lithium in late-type stars}",
      journal = {\mnras},
         year = 2021,
        month = jan,
       volume = {500},
       number = {2},
        pages = {2159-2176},
          doi = {10.1093/mnras/staa3381},
archivePrefix = {arXiv},
       eprint = {2010.15248},
 primaryClass = {astro-ph.SR},
       adsurl = {https://ui.adsabs.harvard.edu/abs/2021MNRAS.500.2159W}
}

@ARTICLE{2017A&A...604A.129M,
       author = {{Mashonkina}, L. and {Jablonka}, P. and {Pakhomov}, Yu. and {Sitnova}, T. and {North}, P.},
        title = "{The formation of the Milky Way halo and its dwarf satellites; a NLTE-1D abundance analysis. I. Homogeneous set of atmospheric parameters}",
      journal = {\aap},
         year = 2017,
        month = aug,
       volume = {604},
          eid = {A129},
        pages = {A129},
          doi = {10.1051/0004-6361/201730779},
archivePrefix = {arXiv},
       eprint = {1704.07656},
 primaryClass = {astro-ph.SR},
       adsurl = {https://ui.adsabs.harvard.edu/abs/2017A&A...604A.129M}
}

\appendix

\section{Results of Program Stars}\label{sec:a1}
\renewcommand{\thetable}{\thesection.\arabic{table}}
\setcounter{table}{0}

\begin{longtable*}{lccrcccc}
   \caption{Li Abundance and Stellar Parameters.} 
   \label{tab:a1}\\
   \hline \multicolumn{1}{c}{Star} 
   & \multicolumn{1}{c}{$T_{\rm{eff}}$(K)}
   & \multicolumn{1}{c}{$\log{g}$(dex)}
   &\multicolumn{1}{c}{[Fe/H]}
   &\multicolumn{1}{c} {$\xi_{\rm t}(km\,s^{-1})$} 
   &\multicolumn{1}{c}{A(Li)$_{\rm{LTE}}$}
   &\multicolumn{1}{c}{A(Li)$_{\rm{NLTE}}$}
   &\multicolumn{1}{c}{phase}
   \\ \hline 
   \endfirsthead
   
   \multicolumn{3}{c}%
   {{\bfseries \tablename\ \thetable{} -- continued from previous page}} \\
   \hline \multicolumn{1}{c}{Star} 
   & \multicolumn{1}{c}{$T_{\rm{eff}}$(K)}
   & \multicolumn{1}{c}{$\log{g}$(dex)}
   &\multicolumn{1}{c}{[Fe/H]}
   &\multicolumn{1}{c} {$\xi_{\rm t}(km\,s^{-1})$} 
   &\multicolumn{1}{c}{A(Li)$_{\rm{LTE}}$}
   &\multicolumn{1}{c}{A(Li)$_{\rm{NLTE}}$}
   &\multicolumn{1}{c}{phase}
   \\ \hline 
   \endhead
   
   \hline \multicolumn{3}{l}{{Continued on next page}} \\ 
   \endfoot
   
   \hline
   \endlastfoot

J$0055+1857$ & 5018 & 2.69 & $-$2.26 & 1.40 & 0.94 & 0.97 & 2  \\ 
J$0119+2425$ & 6412 & 4.27 & $-$2.57 & 1.64 & 2.18 & 2.17 & 1  \\ 
J$0131+4800$ & 6100 & 4.03 & $-$1.79 & 1.35 & 2.21 & 2.23 & 1  \\ 
J$0232+0545$ & 6091 & 3.95 & $-$2.16 & 1.11 & 2.34 & 2.35 & 1  \\ 
J$0244+0828$ & 6472 & 4.39 & $-$2.27 & 1.60 & 2.48 & 2.47 & 1  \\ 
J$0246+2643$ & 6062 & 2.71 & $-$2.38 & 3.30 & 1.44 & \ 1.47\tablenotemark{*} & 4  \\ 
J$0326+0202$ & 4827 & 1.92 & $-$3.05 & 1.52 & 0.85 & 0.88 & 3  \\ 
J$0352+0514$ & 5859 & 4.32 & $-$3.11 & 1.20 & 1.99 & 1.99 & 1  \\ 
J$0423+0538$ & 6100 & 4.12 & $-$2.04 & 1.31 & 2.33 & 2.34 & 1 \\ 
J$0554+5235$ & 5596 & 2.21 & $-$1.93 & 2.30 & 4.04 &\ 3.72\tablenotemark{c} & 4   \\

J$0626+6032$  & 5926 & 3.77 & $-$2.20 & 1.30 & 3.72  & \ 3.54\tablenotemark{c} & 1   \\
J$0637+4308$  & 6366 & 4.11 & $-$2.60 & 1.55 & 2.34  & 2.34 & 1    \\
J$0643+5111$  & 5354 & 3.31 & $-$2.41 & 1.23 & 1.44  & 1.46 & 2   \\
J$0643+5934$  & 4877 & 1.99 & $-$2.46 & 1.51 & 0.99  & 1.03 & 3   \\
J$0705+2552$\tablenotemark{a}  & 5307 & 3.09 & $-$3.11 & 1.42 & 3.94  & \ 3.34\tablenotemark{c} & 2   \\
J$0714+1600$  & 5082 & 2.35 & $-$2.22 & 1.52 & 2.32  & \ 2.34\tablenotemark{c} & 2   \\
J$0748+4613$ & 4645 & 1.40 & $-$3.03 & 2.20 & 0.20 & \ 0.23\tablenotemark{*} & 3  \\ 
J$0828+0037$ & 6104 & 3.48 & $-$2.06 & 1.65 & 2.29 & 2.31 & 1  \\ 
J$0832+2450$ & 6161 & 3.92 & $-$2.25 & 1.50 & 2.20 & 2.20 & 1  \\ 
J$0834+2307$ & 5261 & 3.09 & $-$2.63 & 1.38 & 1.42 & 1.43 & 2  \\ 

J$0845+0150$ & 6310 & 4.01 & $-$2.22 & 1.75 & 2.41 & 2.41 & 1  \\ 
J$0913+0726$ & 6094 & 4.44 & $-$1.93 & 1.34 & 2.36 & 2.36 & 1  \\ 
J$0924+2651$ & 6201 & 3.98 & $-$2.47 & 1.50 & 2.00 & 1.99 & 1  \\ 
J$0934-0108$ & 5045 & 2.25 & $-$2.37 & 1.49 & 1.03 & 1.07 & 2  \\ 
J$0934-0259$ & 5995 & 4.18 & $-$1.82 & 1.30 & 2.24 & 2.22 & 1  \\
J$1014+0547$ & 6051 & 3.75  & $-$1.68 & 1.31  & 2.32  & 2.35  & 1  \\ 
J$1017+3755$ & 5229 & 2.58  & $-$2.78 & 1.51  & 0.87  & 1.27  & 2  \\ 
J$1032+4209$ & 4900 & 1.88  & $-$2.44 & 1.60  & 1.10  & 1.14  & 3 \\
J$1044-0358$\tablenotemark{a}  & 6319 & 4.02 & $-$2.02 & 1.66 & 1.48  & 1.48 & 1   \\
J$1050+2135$  & 6175 & 4.29 & $-$2.26 & 1.27 & 2.27  & 2.25 & 1   \\

J$1058+0138$\tablenotemark{a}  & 6198 & 3.84 & $-$1.97 & 1.26 & 1.58  & 1.59 & 1 \\
J$1059+0208$ & 5072 & 2.74  & $-$2.33 & 1.13  & 1.17  & 1.20  & 2  \\ 
J$1101+2031$ & 5241 & 2.87  & $-$2.48 & 1.45  & 0.54  & 0.57  & 2  \\ 
J$1102+0102$ & 4908 & 2.10  & $-$2.24 & 1.53  & 1.01  & 1.06  & 2  \\ 
J$1109-0754$ & 4582 & 1.09  & $-$3.05 & 1.98  & 0.12  & \ 0.17\tablenotemark{*}  & 3  \\ 
J$1118-0650$ & 4585 & 1.17  & $-$3.31 & 2.14  & $-$0.06  & \ 0.00\tablenotemark{*}  & 3  \\ 
J$1123+0937$ & 5979 & 4.35  & $-$2.49 & 1.33  & 2.19  & 2.18  & 1  \\ 
J$1123+3217$ & 6275 & 4.45  & $-$1.81 & 1.18  & 2.33  & 2.30  & 1  \\ 
J$1135+3100$ & 6289 & 4.22  & $-$2.74 & 1.61  & 2.31  & 2.29  & 1  \\ 
J$1137+4413$ & 6241 & 4.07  & $-$2.08 & 1.52  & 2.42  & 2.43  & 1  \\ 

J$1144+4032$ & 5600 & 4.17  & $-$2.52 & 0.79  & 2.04  & 2.03  & 1  \\ 
J$1147+4458$ & 5668 & 4.36  & $-$2.56 & 0.56  & 1.90  & 1.89  & 1  \\ 
J$1158+0531$ & 5605 & 3.51  & $-$2.23 & 1.39  & 2.12  & 2.14  & 1  \\ 
J$1207+2244$ & 4743 & 1.80  & $-$2.78 & 1.88  & 0.22  & \ 0.25\tablenotemark{*}  & 3  \\ 
J$1210+0023$ & 6033 & 4.38  & $-$2.08 & 1.00  & 2.25  & 2.24  & 1  \\ 
J$1216-0244$ & 6104 & 4.40  & $-$1.95 & 1.06  & 2.43  & 2.43  & 1  \\ 
J$1221+0907$ & 5081 & 2.63  & $-$2.35 & 1.40  & 1.00  & 1.03  & 2  \\ 
J$1225-0452$ & 4872 & 1.90  & $-$2.5 & 1.92  & 1.00  & 1.05  & 3  \\ 
J$1226+2323$ & 4911 & 1.62  & $-$2.39 & 1.94  & 0.26  & \ 0.31\tablenotemark{*}  & 3  \\ 
J$1228+2519$ & 6212 & 4.05  & $-$1.84 & 1.37  & 2.37  & 2.38  & 1  \\ 

J$1231+1232$ & 6259 & 3.93  & $-$2.31 & 0.85  & 2.23  & 2.24  & 1  \\ 
J$1231+5243$ & 5755 & 3.67  & $-$1.96 & 1.38  & 2.32  & 2.34  & 1  \\ 
J$1234+4201$ & 6118 & 4.02  & $-$1.98 & 1.34  & 2.27  & 2.28  & 1  \\ 
J$1237+1922$ & 4928 & 1.65  & $-$2.96 & 1.95  & 0.08  & \ 0.13\tablenotemark{*}  & 3  \\ 
J$1253+0753$ & 6100 & 4.20  & $-$3.68 & 1.08  & 2.07  & 2.07  & 1  \\ 
J$1305+2815$ & 6072 & 3.96  & $-$2.81 & 1.52  & 2.12  & 2.12  & 1 \\
J$1313-0552$\tablenotemark{b}  & 4550 & 1.76 & $-$4.32 & 2.02 & 0.08  & \ 0.15\tablenotemark{*} & 3   \\
J$1345+0513$ & 4614 & 1.17  & $-$2.59 & 2.09  & 1.16  & 1.19  & 3  \\ 
J$1350+0819$ & 6140 & 4.00  & $-$2.33 & 1.70  & 2.34  & 2.34  & 1  \\ 
J$1352+2646$ & 4780 & 1.63  & $-$2.77 & 1.50  & 0.21  & \ 0.23\tablenotemark{*}  & 3  \\ 

J$1353+2021$ & 5959 & 4.25  & $-$2.75 & 1.34  & 2.02  & 2.02  & 1 \\ 
J$1359+2112$\tablenotemark{a}  & 6264 & 3.83 & $-$2.13 & 1.65 & 1.86  & 1.87 & 1   \\
J$1401+2659$  & 6001 & 3.88 & $-$2.89 & 1.50 & 2.03  & 2.04 & 1   \\
J$1404+3222$  & 6252 & 3.81 & $-$1.93 & 1.54 & 2.32  & 2.33 & 1   \\
J$1404+4111$  & 6159 & 3.97 & $-$1.83 & 1.35 & 2.32  & 2.33 & 1   \\
J$1410-0555$\tablenotemark{b}  & 6405 & 4.23 & $-$3.06 & 1.30 & 2.41  & 2.40 & 1   \\
J$1414+1457$ & 4707 & 1.14  & $-$2.50 & 2.01  & 0.23  & \ 0.27\tablenotemark{*}  & 3  \\ 
J$1414+1721$ & 4945 & 2.25  & $-$2.49 & 1.50  & 1.06  & 1.09  & 2  \\ 
J$1423+0322$ & 4540 & 1.05  & $-$2.75 & 2.18  & $-$0.05  & $-$0.02\tablenotemark{*}  & 3  \\
J$1424+3343$ & 5971 & 4.11  & $-$1.88 & 1.25  & 2.22  & 2.24  & 1  \\

J$1432+3755$ & 4585 & 1.34  & $-$3.13 & 2.08  & 0.16  & \ 0.19\tablenotemark{*}  & 3 \\ 
J$1434+3540$\tablenotemark{a}  & 5876 & 4.17 & $-$1.83 & 1.21 & 1.57  & 1.59 & 1   \\
J$1456+3122$ & 4770 & 1.76  & $-$2.78 & 1.63  & 0.74  & 0.77  & 3  \\ 
J$1459+0444$ & 4675 & 1.40  & $-$2.65 & 2.00  & 0.00  & \ 0.03\tablenotemark{*}  & 3  \\ 
J$1518+2544$ & 5417 & 1.90  & $-$2.04 & 2.16  & 0.76  & \ 0.84\tablenotemark{*}  & 4  \\ 
J$1523+0714$ & 4550 & 0.51  & $-$2.57 & 2.07  & 0.10  & \ 0.13\tablenotemark{*}  & 3  \\ 
J$1541+3009$ & 6152 & 4.04  & $-$1.72 & 1.21  & 2.38  & 2.38  & 1  \\ 
J$1548+2113$ & 5140 & 2.75  & $-$2.26 & 1.46  & 1.04  & 1.08  & 2  \\ 
J$1558+4149$ & 6356 & 4.11  & $-$1.83 & 1.33  & 2.48  & 2.49  & 1  \\ 
J$1629+1430$ & 4590 & 1.47  & $-$3.20 & 2.30  & 0.15  & \ 0.18\tablenotemark{*}  & 3  \\ 

J$1634+0206$ & 5247 & 1.91  & $-$2.02 & 1.72  & 0.93  & \ 0.99\tablenotemark{*}  & 4  \\ 
J$1640+1550$ & 4580 & 0.90  & $-$2.89 & 2.34  & 0.18  & \ 0.21\tablenotemark{*}  & 3 \\
J$1657+3443$ & 4770 & 1.60  & $-$2.73 & 1.50  & 0.75  & 0.78  & 3  \\ 
J$1700+2159$ & 5517 & 2.33  & $-$2.48 & 2.06  & 1.07  & \ 1.13\tablenotemark{*}  & 4  \\ 
J$1718+5044$ & 5385 & 2.06  & $-$2.48 & 2.16  & 0.90  & \ 0.96\tablenotemark{*}  & 4  \\ 
J$1730+4143$ & 4700 & 1.36  & $-$2.56 & 1.60  & 0.89  & 0.93  & 3  \\ 
J$1731+2843$ & 5017 & 2.45  & $-$2.09 & 1.40  & 1.11  & 1.16  & 2  \\ 
J$1733+2633$ & 5013 & 2.55  & $-$3.24 & 1.60  & 0.91  & 0.94  & 2  \\ 
J$1829+4852$ & 6213 & 4.16  & $-$1.98 & 1.26  & 2.36  & 2.35  & 1  \\ 
J$1833+1309$ & 5362 & 3.14  & $-$2.45 & 1.15  & 1.60  & 1.62  & 2  \\ 

J$1954+5853$ & 5992 & 3.52  & $-$1.90 & 1.64  & 2.35  & 2.37  & 1 \\ 
J$1958+5533$ & 6286 & 4.02  & $-$2.20 & 1.73  & 2.37  & 2.36  & 1  \\
J$2109+1725$  & 4807 & 1.73 & $-$2.62 & 1.68 & 0.83  & 0.87 & 3   \\
J$2216+0246$  & 4784 & 2.06 & $-$2.33 & 1.90 & 0.64  & 0.68 & 3   \\
J$2216+2232$  & 4681 & 1.40 & $-$2.79 & 1.77 & 0.76  & 0.79 & 3   \\
J$2217+2104$\tablenotemark{b}  & 4507 & 1.21 & $-$3.82 & 2.40 & $-$0.07  & $-$0.01\tablenotemark{*} & 3   \\
J$2221+0228$  & 5219 & 2.86 & $-$2.99 & 1.55 & 1.16  & 1.19 & 2   \\
J$2242+2720$  & 4698 & 1.58 & $-$3.59 & 2.11 & $-$0.01 & \ 0.06\tablenotemark{*} & 3   \\
J$2347+2851$  & 4731 & 1.30 & $-$2.33 & 1.81 & 0.09  & \ 0.12\tablenotemark{*} & 3   \\
J$2350+0236$  & 6171 & 4.18 & $-$2.92 & 1.60 & 2.19  & 2.19 & 1   \\ 

\end{longtable*}
\tablenotetext{}{Note: 1. unevolve stars; 2. RGB stars; 3. highly evolve stars; 4. horizontal branch;}
\tablenotetext{}{\quad\quad\quad a. CEMP-s; b. CEMP-no; c. average of 6707.8 and 6103.6\,\AA;} 
\tablenotetext{}{\quad\quad\quad * . upper limit of A(Li).}


\begin{table*}[hbt]
    \centering
    \caption{Stellar Parameters and Elemental Abundances of Li-rich stars}
    \begin{tabular}{lrrrr}
    \hline
        {Stellar Parameters} & {J0554+5235}&{J0626+6032}&{J0705+2552}&{J0714+1600}\\ \hline
$T_{\rm{eff}}$(K) & 5596 & 5926 & 5307 & 5082 \\ 
$\log{g}$(dex)    & 2.21 & 3.77 & 3.09 & 2.35 \\ 
{[Fe/H]}          & $-$1.93 & $-$2.20 & $-$3.11 &$-$2.22 \\ 
$\xi_{\rm t}(km\,s^{-1})$   & 2.30 & 1.30 & 1.42 & 1.52  \\ 
\hline
Elemental abundances    &   \\  
A(Li)$_{\rm{NLTE}}$     &   3.72 & 3.54 & 3.34 & 2.34   \\ 
{[C/Fe]}$_{\rm{NLTE}}$  &   0.09 & 0.02 & 1.47 & $-$0.09\\ 
{[Na/Fe]}$_{\rm{NLTE}}$ &   0.05 & 0.38 & 1.37 & --     \\
{[Mg/Fe]}$_{\rm{NLTE}}$ &   0.47 & 0.26 & 1.11 & 0.40  \\
{[Ca/Fe]}$_{\rm{NLTE}}$ &   0.36 & --   & 0.60 & --     \\
{[Sc/Fe]}$_{\rm{NLTE}}$ &$-$0.08 & 0.09 & 0.41 & --     \\
{[Ba/Fe]}$_{\rm{NLTE}}$ &$-$0.01 & 0.26 & 0.67 & --     \\
{[Eu/Fe]}$_{\rm{NLTE}}$ &   0.85 & 0.30 & 0.30 & --     \\

        \hline
    \end{tabular}
    \label{tab:5}
\end{table*}


\end{document}